\documentclass[prd,aps,nofootinbib,floatfix,11pt]{revtex4}

\usepackage{amsmath,graphicx,epsfig,amssymb,dsfont,mathtools}
\usepackage[usenames]{color}
\usepackage{ulem} 
\usepackage{bigstrut}
\usepackage{slashed}
\usepackage{multirow}
\usepackage{subfigure}

\allowdisplaybreaks

\begin{document}

\title{Semileptonic $s\to u$ decays of singly heavy baryons}
 
\author{
Yi-Peng Xing$^{1}$, 
Xiao-Qian Jiang$^{1}$, 
Zhen-Xing Zhao$^{1,2}$~\thanks{Email: zhaozx19@imu.edu.cn (Corresponding author)}
}
\affiliation{
$^{1}$ School of Physical Science and Technology, Inner Mongolia University, Hohhot 010021, China\\
 $^{2}$ Research Center for Quantum Physics and Technologies, Inner Mongolia University, Hohhot 010021, China
}
 
\begin{abstract}
$s\to u$ weak decays of heavy flavor hadrons play a unique role in
the field of heavy flavor physics. In this work, we investigate semileptonic
$s\to u$ decays of singly heavy baryons in the light-front approach
under the three-quark picture. Firstly, we provide all the form factors
using this quark model, especially for $f_{3}$ and $g_{3}$, which
are usually considered non-extractable in recent literature. We also
discuss the heavy quark limit for the form factors, and some of our
results differ from those in the literature.
We find that, when considering the heavy quark limit, the $m_{s}$ correction may not be negligible.
Secondly, we apply the obtained form factors to predict some observables in semileptonic
decays. It is worth noting that, due to the
extremely small phase space, lepton flavor universality (LFU) is sensitive
to the masses of leptons, and precise measurement of LFU is expected
to be an important tool for testing the standard model.
\end{abstract}

\maketitle

\section{Introduction}

The study of heavy flavor physics is of great significance for testing
the standard model (SM), searching for new physics, and exploring
the origin of CP violation. In heavy flavor physics, $s\to u$ weak
decays of heavy flavor hadrons play a unique role. On the
one hand, due to the extremely small phase space, lepton flavor universality
(LFU) is sensitive to the masses of leptons, and precise measurement
of LFU is expected to be an important tool for testing the standard
model. On the other hand, the heavy quark symmetry can also be discussed,
and it exhibits properties completely different from the case of
heavy quark decays. 

On the experimental side, there have been many advances in this area.
In 2015, LHCb first discovered evidence of the existence of the decay
channel $\Xi_{b}^{-}\to\Lambda_{b}^{0}\pi^{-}$ \cite{LHCb:2015une},
and in 2023, LHCb confirmed this channel and measured its decay branching
ratio \cite{LHCb:2023tma}. In 2020, LHCb first measured the decay
branching ratio of $\Xi_{c}^{0}\to\Lambda_{c}^{+}\pi^{-}$ \cite{LHCb:2020gge},
and in 2022, Belle revisited this decay channel \cite{Belle:2022kqi},
confirming the previous measurement results of LHCb. These experimental
measurements indicate that, the decay branching ratios of $\Xi_{b}^{-}\to\Lambda_{b}^{0}\pi^{-}$
and $\Xi_{c}^{0}\to\Lambda_{c}^{+}\pi^{-}$ are both of the order
$10^{-3}$. On the theoretical side, there have also been many studies.
Ref. \cite{Faller:2015oma} systematically studied the light-quark
decays of heavy flavor hadrons in the heavy quark limit. Ref. \cite{Soni:2015gwl}
employed the extended harmonic confinement model to understand semileptonic
and pionic decay modes of $\Omega_{Q}$. Ref. \cite{Niu:2021qcc}
investigated the heavy quark conserving weak decays of $\Xi_{Q}$
within the framework of a constituent quark model. Ref. \cite{Liu:2022mxv}
explored the semileptonic and nonleptonic decays of doubly heavy baryons
induced by the $s\to u$ transition in the light-front approach. In
Ref. \cite{Shah:2023qsg}, the authors discussed the semi-electronic
weak decays of the heavy baryons $\Xi_{Q}$ and $\Omega_{Q}$ using
their spectral parameters. Ref. \cite{Shi:2023qnw} considered the semileptonic
decay $B_{s}\to B\ell\nu$ in the covariant light-front approach.
Ref. \cite{Ivanov:2023wir} studied the pionic decay $\Xi_{c}^{0}\to\Lambda_{c}^{+}\pi^{-}$
in the framework of the covariant confined quark model. Other theoretical
works can be found in Refs. \cite{Voloshin:2015xxa,Gronau:2015jgh,Gronau:2016xiq,Soni:2018adu,Ivanov:2019nqd,Voloshin:2019ngb,Cheng:2021qpd,Cheng:2022kea,Cheng:2022jbr}.

In this work, we will adopt the light-front quark model (LFQM) to
investigate semileptonic $s\to u$ decays of singly heavy baryons.
In LFQM, a hadron state is expanded into combinations of constituent
quark states, and the expansion coefficients include color, momentum,
and flavor-spin wavefunctions. LFQM has been successfully applied
to study weak decays of mesons, for a review, readers can refer to
Ref. \cite{Cheng:2003sm}. In recent years, LFQM was further applied
to the baryon sector. In this regard, based on whether a diquark is
explicitly introduced inside a baryon, there are two different pictures
-- the diquark picture and the three-quark picture. In Refs.~\cite{Ke:2007tg,Wei:2009np,Ke:2012wa,Chua:2018lfa,Hu:2020mxk,Hsiao:2020gtc,Wang:2022ias,Liu:2022mxv,Liu:2023zvh},
the diquark picture was adopted, while in Refs.~\cite{Ke:2019smy,Ke:2019lcf,Ke:2021pxk,Geng:2020fng,Geng:2020gjh,Geng:2021nkl,Geng:2022xpn,Xing:2023jnr,Zhao:2023yuk},
the three-quark picture was used. In the diquark picture, the two
spectator quarks are viewed as a ``diquark'', whose spin-parity
can only be $0^{+}$ or $1^{+}$ for an S-wave baryon. However,
the diquark picture has some inherent flaws, including but not limited
to: the two quarks that constitute a diquark can be arbitrarily specified,
and the diquark mass cannot be well determined \cite{Zhao:2023yuk,Zhao:2022vfr}.
In the three-quark picture, the three valence quarks in the baryon are treated independently.
In this work, we will adopt the three-quark picture, and specifically consider the following $s\to u$ transitions:
\begin{itemize}
\item The $0^{+}\to0^{+}$ processes: $\Xi_{c}^{0}(csd)\to\Lambda_{c}^{+}(cud)$
and $\Xi_{b}^{-}(bsd)\to\Lambda_{b}^{0}(bud)$, 
\item The $0^{+}\to1^{+}$ processes: $\Xi_{c}^{+}(csu)\to\Sigma_{c}^{++}(cuu)$,
$\Xi_{c}^{0}(csd)\to\Sigma_{c}^{+}(cud)$, 
\item The $1^{+}\to0^{+}$ processes: $\Omega_{c}^{0}(css)\to\Xi_{c}^{+}(cus)$
and $\Omega_{b}^{-}(bss)\to\Xi_{b}^{0}(bus)$, 
\item The $1^{+}\to1^{+}$ processes: $\Omega_{c}^{0}(css)\to\Xi_{c}^{\prime+}(cus)$
and $\Omega_{b}^{-}(bss)\to\Xi_{b}^{\prime0}(bus)$, 
\end{itemize}
which have been divided into four categories, according to the diquark
types in the initial and final baryons.

The rest of this article is arranged as follows.
In Sec.~II,
we will first introduce the light-front quark model under the three-quark picture,
then elaborate on how to extract the transition form factors,
and finally, discuss in detail the heavy quark limit of these form factors.
In Sec.~III, we will first present the numerical results of form factors,
and then apply them to arrive at some phenomenological predictions.
We conclude this article in the last section.

\section{Light-front approach and transition form factors}

\subsection{Baryons in LFQM}

\label{subsec:The-baryon-states}

In the light-front approach, the light-front momentum is decomposed
into $p=(p^{-},p^{+},p_{\perp})$ with $p^{\pm}\equiv p^{0}\pm p^{3}$
and $p_{\perp}\equiv(p^{1},p^{2})$, and the following notations are
usually adopted 
\begin{equation}
\tilde{p}=(p^{+},p_{\perp}),\quad\{d^{3}\tilde{p}\}=\frac{dp^{+}d^{2}p_{\perp}}{2(2\pi)^{3}}.
\end{equation}

The baryon state in the light-front approach under the three-quark
picture is expressed as 
\begin{eqnarray}
 &  & |{\cal B}(P,S,S_{z})\rangle\nonumber \\
 & = & \int\{d^{3}\tilde{p}_{1}\}\{d^{3}\tilde{p}_{2}\}\{d^{3}\tilde{p}_{3}\}2(2\pi)^{3}\delta^{3}(\tilde{P}-\tilde{p}_{1}-\tilde{p}_{2}-\tilde{p}_{3})\frac{1}{\sqrt{P^{+}}}\nonumber \\
 & \times & \sum_{\lambda_{1},\lambda_{2},\lambda_{3}}\Psi^{SS_{z}}(\tilde{p}_{1},\tilde{p}_{2},\tilde{p}_{3},\lambda_{1},\lambda_{2},\lambda_{3})C^{ijk}|q_{1}^{i}(p_{1},\lambda_{1})q_{2}^{j}(p_{2},\lambda_{2})q_{3}^{k}(p_{3},\lambda_{3})\rangle,\label{eq:baryon_state}
\end{eqnarray}
where $p_{i}$ and $\lambda_{i}$ are respectively the light-front
momentum and helicity of the $i$-th quark, the color wavefunction
$C^{ijk}=\epsilon^{ijk}/\sqrt{6}$, and the flavor-spin and momentum
wavefunctions are contained in $\Psi^{SS_{z}}$. In this work, we
adopt the conventional LFQM, where quarks are all on their mass shells,
so that 
\begin{equation}
p_{i}^{-}=\frac{m_{i}^{2}+p_{i\perp}^{2}}{p_{i}^{+}}.
\end{equation}
The intrinsic variables $(x_{i},k_{i\perp})$ are introduced by 
\begin{align}
 & p_{i}^{+}=x_{i}P^{+},\quad p_{i\perp}=x_{i}P_{\perp}+k_{i\perp},\nonumber \\
 & \sum_{i=1}^{3}x_{i}=1,\quad\sum_{i=1}^{3}k_{i\perp}=0,
\end{align}
where $x_{i}$ is the light-front momentum fraction constrained by
$0\le x_{i}\le1$. Define $\bar{P}\equiv p_{1}+p_{2}+p_{3}$
and $M_{0}^{2}\equiv\bar{P}^{2}$, and then one can show that 
\begin{equation}
M_{0}^{2}=\frac{k_{1\perp}^{2}+m_{1}^{2}}{x_{1}}+\frac{k_{2\perp}^{2}+m_{2}^{2}}{x_{2}}+\frac{k_{3\perp}^{2}+m_{3}^{2}}{x_{3}}.
\end{equation}
The internal momenta $k_{i}$ are defined as 
\begin{align}
k_{i}=(k_{i}^{-},k_{i}^{+},k_{i\perp}) & =(e_{i}-k_{iz},e_{i}+k_{iz},k_{i\perp})\nonumber \\
 & =(\frac{m_{i}^{2}+k_{i\perp}^{2}}{x_{i}M_{0}},x_{i}M_{0},k_{i\perp}),
\end{align}
then it is easy to obtain 
\begin{align}
e_{i} & =\frac{x_{i}M_{0}}{2}+\frac{m_{i}^{2}+k_{i\perp}^{2}}{2x_{i}M_{0}},\nonumber \\
k_{iz} & =\frac{x_{i}M_{0}}{2}-\frac{m_{i}^{2}+k_{i\perp}^{2}}{2x_{i}M_{0}},
\end{align}
where $e_{i}$ is the energy of the $i$-th quark in the rest frame
of $\bar{P}$.

The S-wave singly heavy baryons can be divided into three categories:
$\Lambda_{Q}$-type, $\Sigma_{Q}$-type, and $\Sigma_{Q}^{*}$-type,
with spin $1/2$, $1/2$ and $3/2$, respectively, which correspond
to the following spin coupling decomposition: 
\begin{equation}
\left(\frac{1}{2}\otimes\frac{1}{2}\right)\otimes\frac{1}{2}=(0\oplus1)\otimes\frac{1}{2}=\frac{1}{2}\oplus\frac{1}{2}\oplus\frac{3}{2}.
\end{equation}
For $\Lambda_{Q}$, in which the $u$ and $d$ quarks are considered
to form a scalar diquark, $\Psi$ in Eq. (\ref{eq:baryon_state})
is 
\begin{align}
\Psi_{0}^{S=\frac{1}{2},S_{z}}(\tilde{p}_{i},\lambda_{i})= & A_{0}\bar{u}(p_{3},\lambda_{3})(\bar{\slashed P}+M_{0})(-\gamma_{5})C\bar{u}^{T}(p_{2},\lambda_{2})\nonumber \\
 & \times\bar{u}(p_{1},\lambda_{1})u(\bar{P},S_{z})\Phi(x_{i},k_{i\perp}).\label{eq:wf_LQ}
\end{align}
For $\Sigma_{Q}$, in which the $u$ and $d$ quarks are considered
to form an axial-vector diquark, $\Psi$ is 
\begin{align}
\Psi_{1}^{S=\frac{1}{2},S_{z}}(\tilde{p}_{i},\lambda_{i})= & A_{1}\bar{u}(p_{3},\lambda_{3})(\bar{\slashed P}+M_{0})(\gamma^{\mu}-v^{\mu})C\bar{u}^{T}(p_{2},\lambda_{2})\nonumber \\
 & \times\bar{u}(p_{1},\lambda_{1})(\frac{1}{\sqrt{3}}\gamma_{\mu}\gamma_{5})u(\bar{P},S_{z})\Phi(x_{i},k_{i\perp}),\label{eq:wf_SQ}
\end{align}
where $v^{\mu}\equiv\bar{P}^{\mu}/M_{0}$. The normalization factors
of spin wavefunctions in Eqs. (\ref{eq:wf_LQ}) and (\ref{eq:wf_SQ})
are 
\begin{equation}
A_{0}=A_{1}=\frac{1}{4\sqrt{M_{0}^{3}(e_{1}+m_{1})(e_{2}+m_{2})(e_{3}+m_{3})}}.\label{eq:normalization_factor}
\end{equation}

One comment is in order. If identical quarks are contained in the
baryon, one additional factor should be added. For example, for $\Omega_{c}^{0}(css)$,
when normalizing the baryon state, a factor 2 appears in $\langle{\cal B}(P^{\prime},S^{\prime},S_{z}^{\prime})|{\cal B}(P,S,S_{z})\rangle$
because of two equivalent contractions -- An additional factor $1/\sqrt{2}$
should be added in the definition of $|\Omega_{c}^{0}\rangle$.

The momentum wavefunctions $\Phi$ in Eqs. (\ref{eq:wf_LQ}) and (\ref{eq:wf_SQ})
are normalized by 
\begin{equation}
\int\left(\prod_{i=1}^{3}\frac{dx_{i}d^{2}k_{i\perp}}{2(2\pi)^{3}}\right)2(2\pi)^{3}\delta(1-\sum x_{i})\delta^{2}(\sum k_{i\perp})|\Phi(x_{i},k_{i\perp})|^{2}=1,\label{eq:momentum_wf_normalization}
\end{equation}
and in this work, we adopt the following double Gaussian wavefunction:
\begin{equation}
\Phi(x_{i},k_{i\perp})=\sqrt{\frac{e_{1}e_{2}e_{3}}{x_{1}x_{2}x_{3}M_{0}}}\varphi(\vec{k}_{1},\beta_{1})\varphi(\frac{\vec{k}_{2}-\vec{k}_{3}}{2},\beta_{23}),\label{eq:momentum_wf}
\end{equation}
where $\varphi(\vec{k},\beta)=4\left(\frac{\pi}{\beta^{2}}\right)^{3/4}\exp\left(-\frac{\vec{k}^{2}}{2\beta^{2}}\right)$
with $\vec{k}\equiv(k_{\perp},k_{z})$, and $\beta_{1}$ and $\beta_{23}$
are the shape parameters. Here, we have designated quark
1 as the heavy quark, quark 2 and quark 3 as light quarks; $\beta_{1}$
and $\beta_{23}$ respectively characterize the momentum distribution
between quark 1 and the diquark, and that between light quarks. 

After respectively normalizing the wavefunctions in color, flavor-spin,
and momentum spaces, one can verify that the baryon state is normalized
by 
\begin{equation}
\langle{\cal B}(P^{\prime},S^{\prime},S_{z}^{\prime})|{\cal B}(P,S,S_{z})\rangle=2(2\pi)^{3}P^{+}\delta^{3}(\tilde{P}^{\prime}-\tilde{P})\delta_{S^{\prime}S}\delta_{S_{z}^{\prime}S_{z}}.\label{eq:state_normalization}
\end{equation}

\subsection{Transition form factors}

\label{subsec:FF}

We intend to extract the form factors for four types of processes:
$\Xi_{Q}\to\Lambda_{Q}/\Sigma_{Q}$ and $\Omega_{Q}\to\Xi_{Q}^{(\prime)}$.
There is an important note here, which is about the so-called overlap
factor. For example, when calculating the transition form factors
of $\Xi_{c}^{(\prime)+}(csu)\to\Sigma_{c}^{++}(cuu)$, two multiplication
factors $1/\sqrt{2}\times2$ are present -- the $1/\sqrt{2}$ comes
from the normalization of the flavor wavefunction of $\Sigma_{c}^{++}$,
while the $2$ comes from two equivalent contractions in the transition
matrix element. In the following, we will take $\Xi_{Q}\to\Lambda_{Q}$
as an example, and form factors of other transitions can be obtained
in a similar way. 

On the one hand, the weak transition matrix element $\langle\Lambda_{Q}|\bar{u}\gamma^{\mu}(1-\gamma_{5})s|\Xi_{Q}\rangle$
can be parameterized in terms of the form factors $f_{i}$ and $g_{i}$:
\begin{align}
\langle\Lambda_{Q}(P^{\prime},S_{z}^{\prime})|\bar{u}\gamma^{\mu}s|\Xi_{Q}(P,S_{z})\rangle= & \bar{u}(P^{\prime},S_{z}^{\prime})[\gamma^{\mu}f_{1}(q^{2})+i\sigma^{\mu\nu}\frac{q_{\nu}}{M}f_{2}(q^{2})+\frac{q^{\mu}}{M}f_{3}(q^{2})]u(P,S_{z}),\label{eq:fi_defs}\\
\langle\Lambda_{Q}(P^{\prime},S_{z}^{\prime})|\bar{u}\gamma^{\mu}\gamma_{5}s|\Xi_{Q}(P,S_{z})\rangle= & \bar{u}(P^{\prime},S_{z}^{\prime})[\gamma^{\mu}g_{1}(q^{2})+i\sigma^{\mu\nu}\frac{q_{\nu}}{M}g_{2}(q^{2})+\frac{q^{\mu}}{M}g_{3}(q^{2})]\gamma_{5}u(P,S_{z}),\label{eq:gi_defs}
\end{align}
where $q=P-P^{\prime}$. On the other hand, it can also be calculated
in LFQM 
\begin{align}
 & \langle\Lambda_{Q}(P^{\prime},S_{z}^{\prime})|\bar{u}\gamma^{\mu}(1-\gamma_{5})s|\Xi_{Q}(P,S_{z})\rangle\nonumber \\
= & \int\{d^{3}\tilde{p}_{2}\}\{d^{3}\tilde{p}_{3}\}\frac{A_{0}^{\prime}A_{0}}{\sqrt{p_{1}^{\prime+}p_{1}^{+}P^{\prime+}P^{+}}}\Phi^{\prime*}(x_{i}^{\prime},k_{i\perp}^{\prime})\Phi(x_{i},k_{i\perp})\nonumber \\
\times & {\rm Tr}[(\bar{\slashed P}+M_{0})\gamma_{5}C(\slashed p_{3}+m_{3})^{T}C(-\gamma_{5})(\bar{\slashed P}^{\prime}+M_{0}^{\prime})(\slashed p_{1}^{\prime}+m_{1}^{\prime})\gamma^{\mu}(1-\gamma_{5})(\slashed p_{1}+m_{1})]\nonumber \\
\times & \bar{u}(\bar{P}^{\prime},S_{z}^{\prime})(\slashed p_{2}+m_{2})u(\bar{P},S_{z}).\label{eq:matrix_element_XiQ_LQ}
\end{align}
The form factors $f_{i}$ and $g_{i}$ can be extracted in the following
methods, where we choose the frame that satisfies $q^{+}=0$ or
\begin{equation}
P^{+}=P^{\prime+},
\end{equation}
and make full use of the following key equation (see, for example, Ref. \cite{Chua:2018lfa}): 
\begin{equation}
\gamma^{+}u(P)=\gamma^{+}u(\bar{P}).\label{eq:key_eq}
\end{equation}

Respectively multiply the ``+'' component of Eq.~(\ref{eq:fi_defs})
by $\sum_{S_{z},S_{z}^{\prime}}\bar{u}(P,S_{z})\gamma^{+}u(P^{\prime},S_{z}^{\prime})$
and $\sum_{S_{z},S_{z}^{\prime}}\bar{u}(P,S_{z})(\sum_{j=1}^{2}i\sigma^{+j}q^{j})u(P^{\prime},S_{z}^{\prime})$
from the left to obtain 
\begin{equation}
{\rm Tr}[(\slashed P+M)\gamma^{+}(\slashed P^{\prime}+M^{\prime})(f_{1}\gamma^{+}+f_{2}i\sigma^{+\nu}\frac{q_{\nu}}{M})]=8P^{+}P^{\prime+}f_{1},\label{eq:extract_f1}
\end{equation}
and 
\begin{equation}
{\rm Tr}[(\slashed P+M)(\sum_{j=1}^{2}i\sigma^{+j}q^{j})(\slashed P^{\prime}+M^{\prime})(f_{1}\gamma^{+}+f_{2}i\sigma^{+\nu}\frac{q_{\nu}}{M})]=-8P^{+}P^{\prime+}\frac{q^{2}}{M}f_{2}.
\end{equation}
Respectively multiply the ``+'' component of Eq.~(\ref{eq:gi_defs})
by $\sum_{S_{z},S_{z}^{\prime}}\bar{u}(P,S_{z})\gamma^{+}\gamma_{5}u(P^{\prime},S_{z}^{\prime})$
and $\sum_{S_{z},S_{z}^{\prime}}\bar{u}(P,S_{z})(\sum_{j=1}^{2}i\sigma^{+j}q^{j}\gamma_{5})u(P^{\prime},S_{z}^{\prime})$
from the left to obtain 
\begin{equation}
{\rm Tr}[(\slashed P+M)\gamma^{+}\gamma_{5}(\slashed P^{\prime}+M^{\prime})(g_{1}\gamma^{+}+g_{2}i\sigma^{+\nu}\frac{q_{\nu}}{M})\gamma_{5}]=8P^{+}P^{\prime+}g_{1},
\end{equation}
and 
\begin{equation}
{\rm Tr}[(\slashed P+M)(\sum_{j=1}^{2}i\sigma^{+j}q^{j}\gamma_{5})(\slashed P^{\prime}+M^{\prime})(g_{1}\gamma^{+}+g_{2}i\sigma^{+\nu}\frac{q_{\nu}}{M})\gamma_{5}]=8P^{+}P^{\prime+}\frac{q^{2}}{M}g_{2}.
\end{equation}
Do the same thing to the vector current or axial-vector current part
of Eq. (\ref{eq:matrix_element_XiQ_LQ}), and also note Eq. (\ref{eq:key_eq}),
to obtain 
\begin{align}
f_{1} & =\frac{1}{8P^{+}P^{\prime+}}\int\{d^{3}\tilde{p}_{2}\}\{d^{3}\tilde{p}_{3}\}\frac{A_{0}^{\prime}A_{0}}{\sqrt{p_{1}^{\prime+}p_{1}^{+}P^{\prime+}P^{+}}}\Phi^{\prime*}(x_{i}^{\prime},k_{i\perp}^{\prime})\Phi(x_{i},k_{i\perp})\nonumber \\
 & \times{\rm Tr}[(\bar{\slashed P}+M_{0})\gamma_{5}C(\slashed p_{3}+m_{3})^{T}C(-\gamma_{5})(\bar{\slashed P}^{\prime}+M_{0}^{\prime})(\slashed p_{1}^{\prime}+m_{1}^{\prime})\Gamma_{1}(\slashed p_{1}+m_{1})]\nonumber \\
 & \times{\rm Tr}[(\bar{\slashed P}+M_{0})\Gamma_{2}(\bar{\slashed P}^{\prime}+M_{0}^{\prime})(\slashed p_{2}+m_{2})],\label{eq:expr_ff}
\end{align}
with
\begin{equation}
\Gamma_{1}=\gamma^{+},\quad\Gamma_{2}=\gamma^{+},
\end{equation}
$f_{2}$ with
\begin{equation}
\Gamma_{1}=\gamma^{+},\quad\Gamma_{2}=-\frac{M}{q^{2}}\sum_{j=1}^{2}i\sigma^{+j}q^{j},
\end{equation}
$g_{1}$ with
\begin{equation}
\Gamma_{1}=\gamma^{+}\gamma_{5},\quad\Gamma_{2}=\gamma^{+}\gamma_{5},
\end{equation}
and $g_{2}$ with
\begin{equation}
\Gamma_{1}=\gamma^{+}\gamma_{5},\quad\Gamma_{2}=\frac{M}{q^{2}}\sum_{j=1}^{2}i\sigma^{+j}q^{j}\gamma_{5}.
\end{equation}

The extraction of $f_{3}$ and $g_{3}$ requires some more effort,
where the ``$\perp$'' components of the transition matrix elements
are needed \cite{Cardarelli:1997sx}. We will now elaborate on the extraction
of $f_{3}$ in detail. Similar to the extraction of $f_{1}$, at this
time, multiply the $j$-th ($j=1$ or $2$) component of Eq. (\ref{eq:fi_defs})
by $\sum_{S_{z},S_{z}^{\prime}}\bar{u}(P,S_{z})\gamma^{+}u(P^{\prime},S_{z}^{\prime})$
from the left to obtain
\begin{align}
 & {\rm Tr}[(\slashed P+M)\gamma^{+}(\slashed P^{\prime}+M^{\prime})(f_{1}\gamma^{j}+f_{2}i\sigma^{j\nu}\frac{q_{\nu}}{M}+f_{3}\frac{q^{j}}{M})]\nonumber \\
= & 4P^{+}[f_{1}(2P^{j}-q^{j})+f_{3}q^{j}(1+\frac{M^{\prime}}{M})],\label{eq:f3_had}
\end{align}
where $P^{\prime j}=P^{j}-q^{j}$ has been used. Do the same thing
to the vector current part of Eq. (\ref{eq:matrix_element_XiQ_LQ})
to arrive at
\begin{align}
 & \int\{d^{3}\tilde{p}_{2}\}\{d^{3}\tilde{p}_{3}\}\frac{A_{0}^{\prime}A_{0}}{\sqrt{p_{1}^{\prime+}p_{1}^{+}P^{\prime+}P^{+}}}\Phi^{\prime*}(x_{i}^{\prime},k_{i\perp}^{\prime})\Phi(x_{i},k_{i\perp})\nonumber \\
 & \times{\rm Tr}[(\bar{\slashed P}+M_{0})\gamma_{5}C(\slashed p_{3}+m_{3})^{T}C(-\gamma_{5})(\bar{\slashed P}^{\prime}+M_{0}^{\prime})(\slashed p_{1}^{\prime}+m_{1}^{\prime})\gamma^{j}(\slashed p_{1}+m_{1})]\nonumber \\
 & \times{\rm Tr}[(\bar{\slashed P}+M_{0})\gamma^{+}(\bar{\slashed P}^{\prime}+M_{0}^{\prime})(\slashed p_{2}+m_{2})].\label{eq:f3_LFQM}
\end{align}
One can explicitly check that the $P^{j}$ terms in Eqs. (\ref{eq:f3_had})
and (\ref{eq:f3_LFQM}) precisely cancel each other out. This is not
accidental. On the one hand, when setting ``$j\to+$'' in Eqs. (\ref{eq:f3_had})
and (\ref{eq:f3_LFQM}), we return to the extraction of $f_{1}$.
On the other hand,
\begin{equation}
\bar{P}^{+}=P^{+},\quad\bar{P}^{\prime+}=P^{+},\quad p_{i}^{+}=x_{i}P^{+},\quad p_{1}^{\prime+}=x_{1}P^{+}
\end{equation}
are used for the extraction of $f_{1}$, while
\begin{equation}
\bar{P}_{\perp}=P_{\perp},\quad\bar{P}_{\perp}^{\prime}=P_{\perp}-q_{\perp},\quad p_{i\perp}=x_{i}P_{\perp}+k_{i\perp},\quad p_{1\perp}^{\prime}=x_{1}(P_{\perp}-q_{\perp})+k_{1\perp}^{\prime}
\end{equation}
are used for the extraction of $f_{3}$, from which, one can see that
the coefficients of $P^{+}$ and $P_{\perp}$ are always the same.
In the end, one can obtain 
\begin{align}
 & f_{1}-f_{3}(1+M^{\prime}/M)\nonumber \\
= & \frac{1}{4P^{+}q^{2}}\int\{d^{3}\tilde{p}_{2}\}\{d^{3}\tilde{p}_{3}\}\frac{A_{0}^{\prime}A_{0}}{\sqrt{p_{1}^{\prime+}p_{1}^{+}P^{\prime+}P^{+}}}\Phi^{\prime*}(x_{i}^{\prime},k_{i\perp}^{\prime})\Phi(x_{i},k_{i\perp})\nonumber \\
 & \times\sum_{j=1}^{2}q^{j}\ \{{\rm Tr}[(\bar{\slashed P}+M_{0})\gamma_{5}C(\slashed p_{3}+m_{3})^{T}C(-\gamma_{5})(\bar{\slashed P}^{\prime}+M_{0}^{\prime})(\slashed p_{1}^{\prime}+m_{1}^{\prime})\gamma^{j}(\slashed p_{1}+m_{1})]\nonumber \\
 & \times{\rm Tr}[(\bar{\slashed P}+M_{0})\gamma^{+}(\bar{\slashed P}^{\prime}+M_{0}^{\prime})(\slashed p_{2}+m_{2})]\}|_{P^{j}=0}.\label{eq:for_f3}
\end{align}
In a similar way, one can also arrive at 
\begin{align}
 & g_{1}+g_{3}(1-M^{\prime}/M)\nonumber \\
= & \frac{1}{4P^{+}q^{2}}\int\{d^{3}\tilde{p}_{2}\}\{d^{3}\tilde{p}_{3}\}\frac{A_{0}^{\prime}A_{0}}{\sqrt{p_{1}^{\prime+}p_{1}^{+}P^{\prime+}P^{+}}}\Phi^{\prime*}(x_{i}^{\prime},k_{i\perp}^{\prime})\Phi(x_{i},k_{i\perp})\nonumber \\
 & \times\sum_{j=1}^{2}q^{j}\ \{{\rm Tr}[(\bar{\slashed P}+M_{0})\gamma_{5}C(\slashed p_{3}+m_{3})^{T}C(-\gamma_{5})(\bar{\slashed P}^{\prime}+M_{0}^{\prime})(\slashed p_{1}^{\prime}+m_{1}^{\prime})\gamma^{j}\gamma_{5}(\slashed p_{1}+m_{1})]\nonumber \\
 & \times{\rm Tr}[(\bar{\slashed P}+M_{0})\gamma^{+}\gamma_{5}(\bar{\slashed P}^{\prime}+M_{0}^{\prime})(\slashed p_{2}+m_{2})]\}|_{P^{j}=0}.\label{eq:for_g3}
\end{align}

In practical calculations, the following equations are useful \cite{Zhao:2023yuk}:
\begin{align}
 & x_{2}^{\prime}=x_{2},\quad k_{2\perp}^{\prime}=k_{2\perp}+x_{2}q_{\perp},\nonumber \\
 & x_{3}^{\prime}=x_{3},\quad k_{3\perp}^{\prime}=k_{3\perp}+x_{3}q_{\perp},\nonumber \\
 & x_{1}^{\prime}=x_{1},\quad k_{1\perp}^{\prime}=k_{1\perp}-(1-x_{1})q_{\perp}.\label{eq:xip_kipp}
\end{align}

\subsection{The heavy quark limit}

\label{subsec:HQL}

In order to compare with the predictions in the heavy quark limit,
we also parameterize $\langle\Lambda_{Q}|\bar{u}\gamma^{\mu}(1-\gamma_{5})s|\Xi_{Q}\rangle$
in terms of the form factors
$F_{i}$ and $G_{i}$: 
\begin{align}
\langle\Lambda_{Q}(P^{\prime},S_{z}^{\prime})|\bar{u}\gamma^{\mu}s|\Xi_{Q}(P,S_{z})\rangle= & \bar{u}(P^{\prime},S_{z}^{\prime})[\frac{P^{\mu}}{M}F_{1}(q^{2})+\frac{P^{\prime\mu}}{M^{\prime}}F_{2}(q^{2})+\gamma^{\mu}F_{3}(q^{2})]u(P,S_{z}),\label{eq:Fi_defs}\\
\langle\Lambda_{Q}(P^{\prime},S_{z}^{\prime})|\bar{u}\gamma^{\mu}\gamma_{5}s|\Xi_{Q}(P,S_{z})\rangle= & \bar{u}(P^{\prime},S_{z}^{\prime})[\frac{P^{\mu}}{M}G_{1}(q^{2})+\frac{P^{\prime\mu}}{M^{\prime}}G_{2}(q^{2})+\gamma^{\mu}G_{3}(q^{2})]\gamma_{5}u(P,S_{z}),\label{eq:Gi_defs}
\end{align}
where $M^{(\prime)}$ is the mass of the initial (final) baryon. $F_{i}$
and $G_{i}$ in Eqs. (\ref{eq:Fi_defs}) and (\ref{eq:Gi_defs}) are
related to $f_{i}$ and $g_{i}$ in Eqs. (\ref{eq:fi_defs}) and (\ref{eq:gi_defs})
by
\begin{align}
 & \left(\begin{array}{c}
f_{1}\\
f_{2}\\
f_{3}
\end{array}\right)=\left(\begin{array}{ccc}
\frac{M+M^{\prime}}{2M} & \frac{M+M^{\prime}}{2M^{\prime}} & 1\\
\frac{1}{2} & \frac{M}{2M^{\prime}} & 0\\
\frac{1}{2} & -\frac{M}{2M^{\prime}} & 0
\end{array}\right)\left(\begin{array}{c}
F_{1}\\
F_{2}\\
F_{3}
\end{array}\right),\quad\left(\begin{array}{c}
g_{1}\\
g_{2}\\
g_{3}
\end{array}\right)=\left(\begin{array}{ccc}
\frac{-M+M^{\prime}}{2M} & \frac{-M+M^{\prime}}{2M^{\prime}} & 1\\
\frac{1}{2} & \frac{M}{2M^{\prime}} & 0\\
\frac{1}{2} & -\frac{M}{2M^{\prime}} & 0
\end{array}\right)\left(\begin{array}{c}
G_{1}\\
G_{2}\\
G_{3}
\end{array}\right).\label{eq:FG2fg}
\end{align}
In the heavy quark limit, $M^{\prime}\to M$, thus Eq. (\ref{eq:FG2fg}) reduces into 
\begin{align}
 & \left(\begin{array}{c}
f_{1}\\
f_{2}\\
f_{3}
\end{array}\right)=\left(\begin{array}{ccc}
1 & 1 & 1\\
\frac{1}{2} & \frac{1}{2} & 0\\
\frac{1}{2} & -\frac{1}{2} & 0
\end{array}\right)\left(\begin{array}{c}
F_{1}\\
F_{2}\\
F_{3}
\end{array}\right),\quad\left(\begin{array}{c}
g_{1}\\
g_{2}\\
g_{3}
\end{array}\right)=\left(\begin{array}{ccc}
0 & 0 & 1\\
\frac{1}{2} & \frac{1}{2} & 0\\
\frac{1}{2} & -\frac{1}{2} & 0
\end{array}\right)\left(\begin{array}{c}
G_{1}\\
G_{2}\\
G_{3}
\end{array}\right).\label{eq:FG2fg_HQL}
\end{align}

In this work, we are limited to considering the heavy quark limit
of form factors at $q^{2}=0$. To this end, we take $m_{u}=m_{d}=m_{s}=0$,
and then one can easily check that, at the leading order of $m_{Q}$,
\begin{equation}
M_{0}^{\prime}=M_{0},\quad e_{i}^{\prime}=e_{i},\;i=1,2,3.
\end{equation}
Furthermore, the normalization factors of the initial and final spin
wavefunctions are equal:
\begin{equation}
A_{0}^{\prime}=A_{0}.
\end{equation}
The following explicit expressions at the leading order of the heavy
quark limit are sometimes useful:
\begin{align}
 & M_{0}=\frac{1}{\sqrt{x_{2}}}m_{2},\quad e_{1}=\frac{x_{1}}{2\sqrt{x_{2}}}m_{2},\quad e_{2}=\frac{1+x_{2}}{2\sqrt{x_{2}}}m_{2},\quad e_{3}=\frac{x_{3}}{2\sqrt{x_{2}}}m_{2},\label{eq:HQL_M0ei}\\
 & x_{1}=0,\quad x_{2}=1,\quad x_{3}=0,
\end{align}
where quark 2 is the heavy quark.

Finally, by comparing the expressions of the form factors with the
normalization of the baryon state, $\langle{\cal B}(P^{\prime},S_{z}^{\prime})|{\cal B}(P,S_{z})\rangle$,
one can see that, the extremely complicated form factors, when evaluated
at $q^{2}=0$ and taking the heavy quark limit, degenerate into the
simple expressions in Table \ref{Tab:ff_HQL}. Some comments are in order. 

\begin{figure}
\includegraphics[scale=0.7]{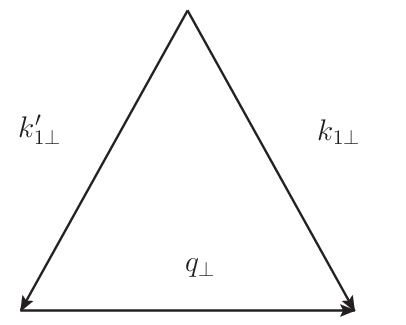}
\caption{
An isosceles triangle illustrating $k_{1\perp}^{\prime}\cdot q_{\perp}=-k_{1\perp}\cdot q_{\perp}$.
}
\label{fig:isosceles}
\end{figure}

\begin{itemize}
\item The form factor $f_{3}$ contains the factor $M/(M+M^{\prime})\approx1/2$,
while the form factor $g_{3}$ contains a large factor $M/(M-M^{\prime})$,
which can be seen from Eqs. (\ref{eq:for_f3}) and (\ref{eq:for_g3}).

\item In Table \ref{Tab:ff_HQL}, the term $(k_{1\perp}+k_{1\perp}^{\prime})\cdot q_{\perp}/q_{\perp}^{2}$
appears many times. At first glance, we may think that this term could
be infinite when we take the limit $q^{2}=-q_{\perp}^{2}\to0$. In
fact, this item tends to 0, and the reason is given as follows. From
Eq. (\ref{eq:xip_kipp}), when taking the limit $x_{1}\to0$, we have
$k_{1\perp}^{\prime}=k_{1\perp}-q_{\perp}$. Considering that, in the heavy
quark limit, the initial $s$ quark (quark $1$) and the final $u$
quark (quark $1^{\prime}$) are on an equal footing, we should have
$|k_{1\perp}|=|k_{1\perp}^{\prime}|$, that is, $k_{1\perp}$, $k_{1\perp}^{\prime}$
and $q_{\perp}$ form an isosceles triangle, as shown in Fig. \ref{fig:isosceles}. At
this point, one can easily see that $k_{1\perp}^{\prime}\cdot q_{\perp}=-k_{1\perp}\cdot q_{\perp}$,
so that $(k_{1\perp}+k_{1\perp}^{\prime})\cdot q_{\perp}/q_{\perp}^{2}=0$.
However, in practical numerical calculations, when taking a small
$|q^{2}|$, we always arrive at a large result with a larger calculation error,
see Table \ref{Tab:f0g0_with_errs} below. This may not be difficult to understand -- for
$(k_{1\perp}+k_{1\perp}^{\prime})\cdot q_{\perp}/q_{\perp}^{2}$ ,
a slight deviation of $(k_{1\perp}+k_{1\perp}^{\prime})\cdot q_{\perp}$
from $0$ may be amplified by $1/q_{\perp}^{2}$.

\item In Table \ref{Tab:ff_HQL}, $1/x_{1}$ also appears many times, where
$x_{1}$ is the momentum fraction of quark 1. From Eq. (\ref{eq:HQL_M0ei}),
we have $1/x_{1}=M_{0}/(2e_{1})$. Considering $M_{0}=m_{2}$ in the
heavy quark limit and $e_{1}\sim\Lambda_{{\rm QCD}}$, we conclude
that $1/x_{1}$ is of order $m_{Q}$. Therefore, in Table \ref{Tab:ff_HQL},
the $f_{2}$ for the processes of $0^{+}\to1^{+}$, $1^{+}\to0^{+}$, and $1^{+}\to1^{+}$
are all of order $m_{Q}$. Combined with the
last comment, one can see that, other form factors are all finite,
or, at most order 1. In the following text,
we will use this conclusion to derive the heavy quark limit of the semi-electronic decay width.
\end{itemize}

\begin{table}
\caption{
Our predictions for the form factors at $q^{2}=0$ in the heavy quark
limit, where the integration in Eq. (\ref{eq:momentum_wf_normalization})
is understood. Note that, we do not consider the overlap factors here.
The term $(k_{1\perp}+k_{1\perp}^{\prime})\cdot q_{\perp}/q_{\perp}^{2}$
should tend to 0 in the heavy quark limit,
however, in practical calculations, significant uncertainties are introduced,
see the explanation in the main text and compare with the numerical results in Table \ref{Tab:f0g0_with_errs}. 
}
\label{Tab:ff_HQL}%
\begin{tabular}{c|c|c|c|c|c|c}
\hline 
 & $f_{1}(0)$ & $f_{2}(0)$ & $f_{3}(0)$ & $g_{1}(0)$ & $g_{2}(0)$ & $g_{3}(0)$\tabularnewline
\hline 
$0^{+}\to0^{+}$ & $1$ & $\frac{1}{2}$ & $\frac{1}{2}\frac{1}{x_{1}}\frac{(k_{1\perp}+k_{1\perp}^{\prime})\cdot q_{\perp}}{q_{\perp}^{2}}$ & $0$ & $0$ & $0$\tabularnewline
$0^{+}\to1^{+}$ & $0$ & $\frac{1}{\sqrt{3}}\frac{1}{x_{1}}$ & $0$ & $-\frac{1}{\sqrt{3}}$ & $-\frac{1}{\sqrt{3}}(\frac{1}{2}+\frac{1}{x_{1}}\frac{(k_{1\perp}+k_{1\perp}^{\prime})\cdot q_{\perp}}{q_{\perp}^{2}})$ & $\frac{1}{\sqrt{3}}\frac{M}{M-M^{\prime}}\frac{1}{x_{1}}\frac{(k_{1\perp}+k_{1\perp}^{\prime})\cdot q_{\perp}}{q_{\perp}^{2}}$\tabularnewline
$1^{+}\to0^{+}$ & $0$ & $\frac{1}{\sqrt{3}}\frac{1}{x_{1}}$ & $0$ & $-\frac{1}{\sqrt{3}}$ & $-\frac{1}{\sqrt{3}}(-\frac{1}{2}+\frac{1}{x_{1}}\frac{(k_{1\perp}+k_{1\perp}^{\prime})\cdot q_{\perp}}{q_{\perp}^{2}})$ & $\frac{1}{\sqrt{3}}\frac{M}{M-M^{\prime}}\frac{1}{x_{1}}\frac{(k_{1\perp}+k_{1\perp}^{\prime})\cdot q_{\perp}}{q_{\perp}^{2}}$\tabularnewline
$1^{+}\to1^{+}$ & $1$ & $-\frac{2}{3}\frac{1}{x_{1}}$ & $\frac{1}{2}\frac{1}{x_{1}}\frac{(k_{1\perp}+k_{1\perp}^{\prime})\cdot q_{\perp}}{q_{\perp}^{2}}$ & $\frac{2}{3}$ & $\frac{2}{3}\frac{1}{x_{1}}\frac{(k_{1\perp}+k_{1\perp}^{\prime})\cdot q_{\perp}}{q_{\perp}^{2}}$ & $-\frac{2}{3}\frac{M}{M-M^{\prime}}\frac{1}{x_{1}}\frac{(k_{1\perp}+k_{1\perp}^{\prime})\cdot q_{\perp}}{q_{\perp}^{2}}$\tabularnewline
\hline 
\end{tabular}
\end{table}

After figuring out these form factors $f_{i}$ and $g_{i}$ at $q^{2}=0$
in the heavy quark limit, we convert them into $F_{i}$ and $G_{i}$
using Eq. (\ref{eq:FG2fg_HQL}). Then, we will compare our results
with those in Ref. \cite{Faller:2015oma}. Although here we consider $q^{2}=0$
and Ref. \cite{Faller:2015oma} considers $q^{2}=q_{{\rm max}}^{2}$, we still
believe that the comparison is reasonable, considering that the phase
space is small and forms factors should be slowly changing functions.
The specific comparisons are given below.

\begin{itemize}
\item For the processes of $0^{+}\to0^{+}$, we have
\begin{align}
\langle\Lambda_{Q}(v^{\prime},s^{\prime})|\bar{u}\gamma_{\mu}s|\Xi_{Q}(v,s)\rangle & =\frac{1}{2}\bar{u}_{\Lambda}(v^{\prime},s^{\prime})u_{\Xi}(v,s)(v+v^{\prime})_{\mu}=\bar{u}_{\Lambda}(v^{\prime},s^{\prime})u_{\Xi}(v,s)v_{\mu},\nonumber \\
\langle\Lambda_{Q}(v^{\prime},s^{\prime})|\bar{u}\gamma_{\mu}\gamma_{5}s|\Xi_{Q}(v,s)\rangle & =0,
\end{align}
 while in Ref. \cite{Faller:2015oma}, 
\begin{align}
\langle\Lambda_{Q}(v^{\prime},s^{\prime})|\bar{u}\gamma_{\mu}s|\Xi_{Q}(v,s)\rangle & =\bar{u}_{\Lambda}(v^{\prime},s^{\prime})u_{\Xi}(v,s)(v+v^{\prime})_{\mu}=2\bar{u}_{\Lambda}(v^{\prime},s^{\prime})u_{\Xi}(v,s)v_{\mu},\nonumber \\
\langle\Lambda_{Q}(v^{\prime},s^{\prime})|\bar{u}\gamma_{\mu}\gamma_{5}s|\Xi_{Q}(v,s)\rangle & =0.
\end{align}

\item For the processes of $1^{+}\to1^{+}$, we have
\begin{align}
\langle\Xi_{Q}^{\prime}(v^{\prime},s^{\prime})|\bar{u}\gamma_{\mu}s|\Omega_{Q}(v,s)\rangle & =\sqrt{2}\bar{u}_{\Xi^{\prime}}(v^{\prime},s^{\prime})[(-\frac{2}{3x_{1}})v_{\mu}+(-\frac{2}{3x_{1}})v_{\mu}^{\prime}+(1+\frac{4}{3x_{1}})\gamma_{\mu}]u_{\Omega}(v,s),\nonumber \\
 & =\sqrt{2}\bar{u}_{\Xi^{\prime}}(v^{\prime},s^{\prime})u_{\Omega}(v,s)v_{\mu},\nonumber \\
\langle\Xi_{Q}^{\prime}(v^{\prime},s^{\prime})|\bar{u}\gamma_{\mu}\gamma_{5}s|\Omega_{Q}(v,s)\rangle & =\sqrt{2}\cdot\frac{2}{3}\bar{u}_{\Xi^{\prime}}(v^{\prime},s^{\prime})\gamma_{\mu}\gamma_{5}u_{\Omega}(v,s)=0,
\end{align}
where $\sqrt{2}$ is the overlap factor, while in Ref. \cite{Faller:2015oma} (see
Eq. (18) therein), 
\begin{align}
\langle\Xi_{Q}^{\prime}(v^{\prime},s^{\prime})|\bar{u}\gamma_{\mu}s|\Omega_{Q}(v,s)\rangle & =-\bar{u}_{\Xi^{\prime}}(v^{\prime},s^{\prime})u_{\Omega}(v,s)(v+v^{\prime})_{\mu}=-2\bar{u}_{\Xi^{\prime}}(v^{\prime},s^{\prime})u_{\Omega}(v,s)v_{\mu},\nonumber \\
\langle\Xi_{Q}^{\prime}(v^{\prime},s^{\prime})|\bar{u}\gamma_{\mu}\gamma_{5}s|\Omega_{Q}(v,s)\rangle & =0.
\end{align}

\item For the processes of $0^{+}\to1^{+}$, we have
\begin{align}
\langle\Sigma_{Q}(v^{\prime},s^{\prime})|\bar{u}\gamma_{\mu}s|\Xi_{Q}(v,s)\rangle & =\frac{1}{\sqrt{3}x_{1}}\bar{u}_{\Sigma}(v^{\prime},s^{\prime})(v_{\mu}+v_{\mu}^{\prime}-2\gamma_{\mu})u_{\Xi}(v,s)=0,\nonumber \\
\langle\Sigma_{Q}(v^{\prime},s^{\prime})|\bar{u}\gamma_{\mu}\gamma_{5}s|\Xi_{Q}(v,s)\rangle & =-\frac{1}{2\sqrt{3}}\bar{u}_{\Sigma}(v^{\prime},s^{\prime})(v_{\mu}+v_{\mu}^{\prime}+2\gamma_{\mu})\gamma_{5}u_{\Xi}(v,s)\nonumber \\
 & =-\frac{1}{\sqrt{3}}\bar{u}_{\Sigma}(v^{\prime},s^{\prime})(v_{\mu}^{\prime}+\gamma_{\mu})\gamma_{5}u_{\Xi}(v,s),
\end{align}
where we have inferred that the vector current matrix element is zero
at the leading order of the heavy quark limit, while in Ref. \cite{Faller:2015oma},
\begin{align}
\langle\Sigma_{Q}(v^{\prime},s^{\prime})|\bar{u}\gamma_{\mu}s|\Xi_{Q}(v,s)\rangle & =0,\nonumber \\
\langle\Sigma_{Q}(v^{\prime},s^{\prime})|\bar{u}\gamma_{\mu}\gamma_{5}s|\Xi_{Q}(v,s)\rangle & =-\frac{1}{\sqrt{3}}\bar{u}_{\Sigma}(v^{\prime},s^{\prime})(v_{\mu}^{\prime}-\gamma_{\mu})\gamma_{5}u_{\Xi}(v,s).
\end{align}

\item For the processes of $1^{+}\to0^{+}$, we have
\begin{align}
\langle\Xi_{Q}(v^{\prime},s^{\prime})|\bar{u}\gamma_{\mu}s|\Omega_{Q}(v,s)\rangle & =\sqrt{2}\cdot\frac{1}{\sqrt{3}x_{1}}\bar{u}_{\Xi}(v^{\prime},s^{\prime})(v_{\mu}+v_{\mu}^{\prime}-2\gamma_{\mu})u_{\Omega}(v,s)=0,\nonumber \\
\langle\Xi_{Q}(v^{\prime},s^{\prime})|\bar{u}\gamma_{\mu}\gamma_{5}s|\Omega_{Q}(v,s)\rangle & =\sqrt{2}\cdot\frac{1}{2\sqrt{3}}\bar{u}_{\Xi}(v^{\prime},s^{\prime})(v_{\mu}+v_{\mu}^{\prime}-2\gamma_{\mu})\gamma_{5}u_{\Omega}(v,s)\nonumber \\
 & =\sqrt{2}\cdot\frac{1}{\sqrt{3}}\bar{u}_{\Xi}(v^{\prime},s^{\prime})(v_{\mu}-\gamma_{\mu})\gamma_{5}u_{\Omega}(v,s),
\end{align}
where $\sqrt{2}$ is the overlap factor, while in Ref. \cite{Faller:2015oma} (same
as those in $0^{+}\to1^{+}$), 
\begin{align}
\langle\Xi_{Q}(v^{\prime},s^{\prime})|\bar{u}\gamma_{\mu}s|\Omega_{Q}(v,s)\rangle & =0,\nonumber \\
\langle\Xi_{Q}(v^{\prime},s^{\prime})|\bar{u}\gamma_{\mu}\gamma_{5}s|\Omega_{Q}(v,s)\rangle & =-\frac{1}{\sqrt{3}}\bar{u}_{\Xi}(v^{\prime},s^{\prime})(v_{\mu}-\gamma_{\mu})\gamma_{5}u_{\Omega}(v,s).
\end{align}
\end{itemize}

When deriving our results above, we have used the following conclusions.
In the limit $P^{\prime}\to P$, the following equations strictly
hold: 
\begin{align}
\bar{u}(P,S_{z}^{\prime})\gamma^{\mu}u(P,S_{z}) & =\bar{u}(P,S_{z}^{\prime})v^{\mu}u(P,S_{z}),\label{eq:vector_bi_HQL}\\
\bar{u}(P,S_{z}^{\prime})\gamma^{\mu}\gamma_{5}u(P,S_{z}) & =\bar{u}(P,S_{z}^{\prime})v^{\mu}\gamma_{5}u(P,S_{z})=0,\label{eq:axial_bi_HQL}
\end{align}
where $v^{\mu}=P^{\mu}/M$, $S_{z}^{\prime}$ and $S_{z}$ do not
need to be the same. These equations can be easily checked in the
rest frame of $P$. Eq. (\ref{eq:vector_bi_HQL}) implies that the
spin of the initial and final states is conserved in the heavy quark
limit. For the $0^{+}\to0^{+}$ case, it is obvious because $S_{z}^{\prime}=s_{Q}=S_{z}$
with $s_{Q}$ the spin of the heavy quark, while for the other three
cases, it is not obvious. Note that, our discussion in this paragraph
is model-independent.

\section{Numerical results and phenomenological applications}

In this section, we will present the numerical results of the form
factors, and then apply them to study semileptonic decays. 

\subsection{Inputs}

The following quark masses are adopted (in units of GeV): 
\begin{equation}
m_{u}=m_{d}=0.25,\quad m_{s}=0.37,\quad m_{c}=1.4,\quad m_{b}=4.8.\label{eq:quark_masses}
\end{equation}
Similar parameters can also be found in, for example, Refs. \cite{Lu:2007sg,Wang:2007sxa,Wang:2008xt,Wang:2008ci,Wang:2009mi,Chen:2009qk,Li:2010bb}.
When estimating the uncertainties introduced by these quark masses,
we consider the following ranges (in units of GeV): 
\begin{equation}
m_{u}=m_{d}=0.22\text{-}0.28,\quad m_{s}=0.37\text{-}0.5,\quad m_{c}=1.3\text{-}1.81,\quad m_{b}=4.7\text{-}5.2.\label{eq:quark_mass_ranges}
\end{equation}

For the shape parameters, we adopt the following values (in units
of GeV):
\begin{align}
 & \beta_{b[ud]}=0.66,\quad\beta_{c[ud]}=0.56,\quad\beta_{[ud]}=0.32,\nonumber \\
 & \beta_{b[sq]}=0.68,\quad\beta_{c[sq]}=0.58,\quad\beta_{[sq]}=0.39,\nonumber \\
 & \beta_{b\{qq\}}=0.68,\quad\beta_{c\{qq\}}=0.58,\quad\beta_{\{qq\}}=0.39,\nonumber \\
 & \beta_{b\{sq\}}=0.73,\quad\beta_{c\{sq\}}=0.63,\quad\beta_{\{sq\}}=0.42,\nonumber \\
 & \beta_{b\{ss\}}=0.78,\quad\beta_{c\{ss\}}=0.66,\quad\beta_{\{ss\}}=0.44,\label{eq:beta_values}
\end{align}
where $q=u/d$, and $[q_{1}q_{2}]$ and $\{q_{1}q_{2}\}$ respectively
denote a scalar diquark and an axial-vector diquark. Some comments
are in order. 
\begin{itemize}
\item The shape parameter between a bottom (charm) quark and a diquark $\beta_{b(c),di}$
is taken between $\beta_{b\bar{s}}=0.623$ GeV ($\beta_{c\bar{s}}=0.535$
GeV) and $\beta_{b\bar{c}}=0.886$ GeV ($\beta_{c\bar{c}}=0.753$
GeV). Here the shape parameters for mesons are determined using the
method proposed in Ref. \cite{Cheng:2003sm}, with the quark masses in Eq.
(\ref{eq:quark_masses}) as inputs. 
\item $\beta_{Q\{qq\}}$ should be slightly larger than $\beta_{Q[ud]}$,
and may be close to $\beta_{Q[sq]}$, where $Q=b/c$. 
\item For the shape parameters of diquarks, we adopt these approximations:
$\beta_{[ud]}=\beta_{u\bar{d}}$, $\beta_{[sq]}=\beta_{s\bar{u}}$
and $\beta_{\{ss\}}=\beta_{s\bar{s}}$.
\item It is expected that there exists approximately 10\% uncertainty in
these shape parameters for baryons. 
\end{itemize}

The masses of initial and final states are collected in Table \ref{Tab:masses}
\cite{ParticleDataGroup:2022pth}.

\begin{table}
\caption{
Masses (in units of MeV) of initial and final baryons. For the masses
of electron and muon, we respectively take $m_{e}=0.511$ MeV and
$m_{\mu}=105.658$ MeV, whose uncertainties can be neglected.
}
\label{Tab:masses}%
\begin{tabular}{l|c|c|c}
\hline 
Transition & $M_{i}$ & $M_{f}$ & $M_{i}-M_{f}$\tabularnewline
\hline 
$\Xi_{c}^{0}\to\Lambda_{c}^{+}$ & $2470.44\pm0.28$ & $2286.46\pm0.14$ & $183.98\pm0.42$\tabularnewline
$\Xi_{b}^{-}\to\Lambda_{b}^{0}$ & $5797.0\pm0.6$ & $5619.6\pm0.17$ & $177.4\pm0.77$\tabularnewline
\hline 
$\Xi_{c}^{+}\to\Sigma_{c}^{++}$ & $2467.71\pm0.23$ & $2453.97\pm0.14$ & $13.74\pm0.37$\tabularnewline
$\Xi_{c}^{0}\to\Sigma_{c}^{+}$ & $2470.44\pm0.28$ & $2452.65_{-0.16}^{+0.22}$ & $17.79_{-0.50}^{+0.44}$\tabularnewline
\hline 
$\Omega_{c}^{0}\to\Xi_{c}^{+}$ & $2695.2\pm1.7$ & $2467.71\pm0.23$ & $227.49\pm1.93$\tabularnewline
$\Omega_{b}^{-}\to\Xi_{b}^{0}$ & $6045.8\pm0.8$ & $5791.9\pm0.5$ & $253.9\pm1.3$\tabularnewline
\hline 
$\Omega_{c}^{0}\to\Xi_{c}^{\prime+}$ & $2695.2\pm1.7$ & $2578.2\pm0.5$ & $117.0\pm2.2$\tabularnewline
$\Omega_{b}^{-}\to\Xi_{b}^{\prime0}$ & $6045.8\pm0.8$ & $5935.1\pm0.5$ & $110.7\pm1.3$\tabularnewline
\hline 
\end{tabular}
\end{table}

\subsection{Form factors and uncertainties}

One can see from Eqs. (\ref{eq:expr_ff}), (\ref{eq:for_f3}) and
(\ref{eq:for_g3}) that, $f_{2,3}$ and $g_{2,3}$ are all proportional
to $1/q^{2}$. In the limit of $q^{2}\to0$, they are all indeterminate
forms of the type $0/0$. Referring to the results in the heavy quark
limit in Table \ref{Tab:ff_HQL}, one can see that some form factors
have definite limits, while some of them contain the factor $(k_{1\perp}+k_{1\perp}^{\prime})\cdot q_{\perp}/q_{\perp}^{2}$,
which introduces significant calculation errors, and the reason has
been given in the previous text. For an illustration, readers can
refer to Table \ref{Tab:f0g0_with_errs}. In light of this, in this
work, we also take into account the calculation errors. 

\begin{table}
\caption{
Our calculated form factors at $q^{2}=0$, where the calculation errors
are also shown. $g_{3}$ is proportional to $M/(M-M^{\prime})$,
so that some of them have particularly large central values and calculation errors.
}
\label{Tab:f0g0_with_errs}%
\begin{tabular}{l|r|r|r|r|r|r}
\hline 
Transition & $f_{1}(0)$ & $f_{2}(0)$ & $f_{3}(0)$ & $g_{1}(0)$ & $g_{2}(0)$ & $g_{3}(0)$\tabularnewline
\hline 
$\Xi_{c}^{0}\to\Lambda_{c}^{+}$ & $0.963\pm0.001$ & $0.419\pm0.002$ & $-0.179\pm0.807$ & $0.000\pm0.000$ & $0.001\pm0.000$ & $-0.025\pm0.001$\tabularnewline
$\Xi_{b}^{-}\to\Lambda_{b}^{0}$ & $0.961\pm0.001$ & $0.447\pm0.001$ & $-0.625\pm2.943$ & $0.000\pm0.000$ & $0.000\pm0.000$ & $-0.027\pm0.001$\tabularnewline
\hline 
$\Xi_{c}^{0}\to\Sigma_{c}^{+}$ & $-0.000\pm0.000$ & $1.311\pm0.027$ & $-0.020\pm0.000$ & $-0.345\pm0.000$ & $0.146\pm0.254$ & $50.692\pm75.260$\tabularnewline
\hline 
$\Omega_{c}^{0}\to\Xi_{c}^{+}$ & $-0.000\pm0.000$ & $1.921\pm0.041$ & $-0.027\pm0.000$ & $-0.463\pm0.000$ & $0.012\pm0.392$ & $3.342\pm10.751$\tabularnewline
$\Omega_{b}^{-}\to\Xi_{b}^{0}$ & $-0.000\pm0.000$ & $4.242\pm0.128$ & $-0.027\pm0.000$ & $-0.444\pm0.001$ & $0.072\pm1.349$ & $13.070\pm80.887$\tabularnewline
\hline 
$\Omega_{c}^{0}\to\Xi_{c}^{\prime+}$ & $1.396\pm0.001$ & $-1.599\pm0.045$ & $-0.533\pm1.468$ & $0.531\pm0.000$ & $-0.131\pm0.471$ & $-11.406\pm26.584$\tabularnewline
$\Omega_{b}^{-}\to\Xi_{b}^{\prime0}$ & $1.392\pm0.002$ & $-4.219\pm0.156$ & $-1.374\pm4.838$ & $0.511\pm0.001$ & $-0.337\pm1.605$ & $-68.078\pm219.844$\tabularnewline
\hline 
\end{tabular}
\end{table}

In addition to the calculation errors mentioned above, we also consider
errors from other sources: (1) quark masses, (2) shape parameters,
and (3) masses of initial and final baryons.

The form factors $f_{i}$ and $g_{i}$ are calculated in the spacelike
region of $q^{2}$. To extrapolate to the physical region of $q^{2}$,
we adopt the following fit formula: 
\begin{equation}
F(q^{2})=\frac{a+b\;z(q^{2})}{1-q^{2}/(m_{{\rm pole}}^{f})^{2}},
\end{equation}
where 
\begin{equation}
z(q^{2})=\frac{\sqrt{t_{+}-q^{2}}-\sqrt{t_{+}-t_{0}}}{\sqrt{t_{+}-q^{2}}+\sqrt{t_{+}-t_{0}}}
\end{equation}
with $t_{+}=(m_{K}+m_{\pi})^{2}$ and $t_{0}=q_{{\rm max}}^{2}=(M-M^{\prime})^{2}$.
The pole masses $m_{{\rm pole}}^{f}$ are respectively taken as $m_{{\rm pole}}^{f_{1},f_{2}}=m_{K^{*}}$,
$m_{{\rm pole}}^{f_{3}}=m_{K_{0}^{*}(700)}$, $m_{{\rm pole}}^{g_{1},g_{2}}=m_{K_{1}(1270)}$,
and $m_{{\rm pole}}^{g_{3}}=m_{K}$. 

\begin{figure}[!htbp]
\centering
\subfigure[\hspace{0.5em}$\Xi_{c}^{0}\to\Lambda_{c}^{+}$]{\includegraphics[width=0.48\textwidth]{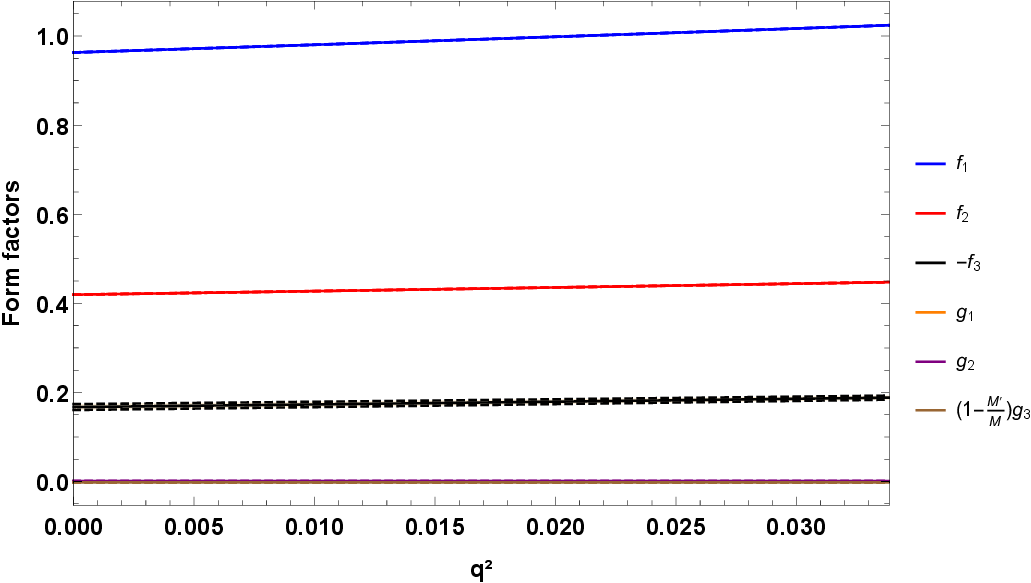}}
    \hfill
    \subfigure[\hspace{0.5em}$\Xi_{b}^{-}\to\Lambda_{b}^{0}$]{\includegraphics[width=0.48\textwidth]{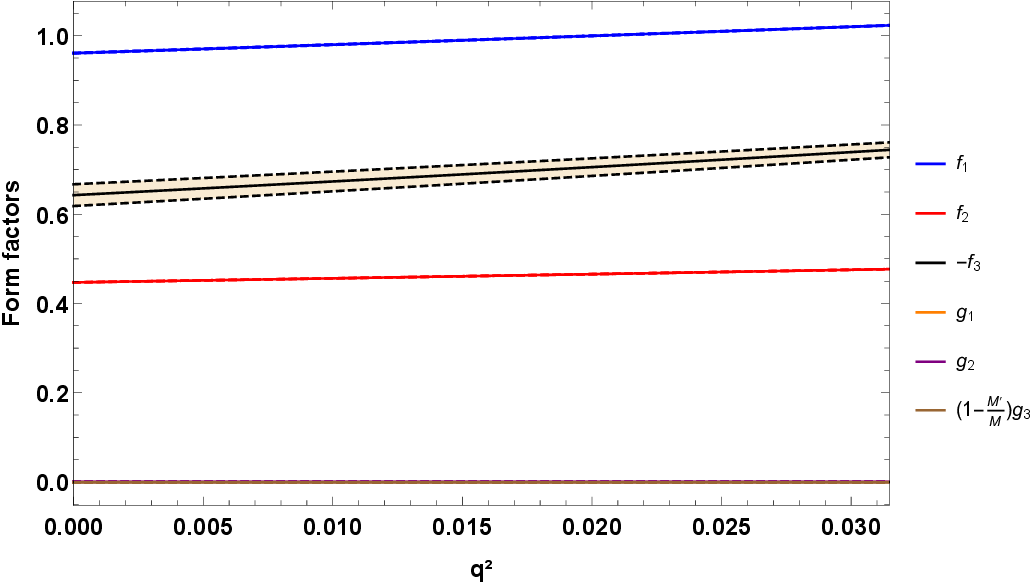}}

    \centering
    \subfigure[\hspace{0.5em}$\Xi_{c}^{0}\to\Sigma_{c}^{+}$]{\includegraphics[width=0.5\textwidth]{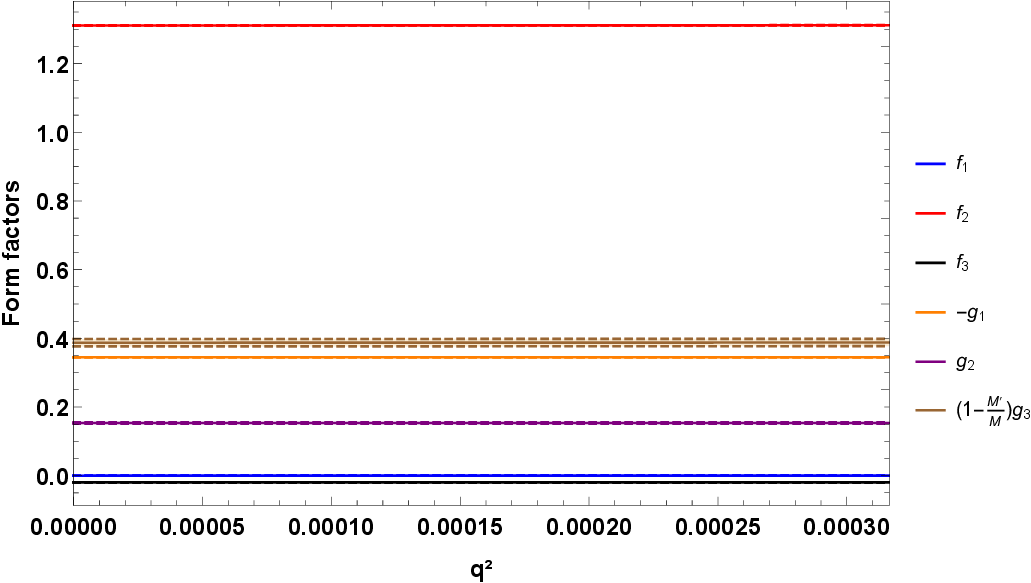}}

    \centering
    \subfigure[\hspace{0.5em}$\Omega_{c}^{0}\to\Xi_{c}^{+}$]{\includegraphics[width=0.48\textwidth]{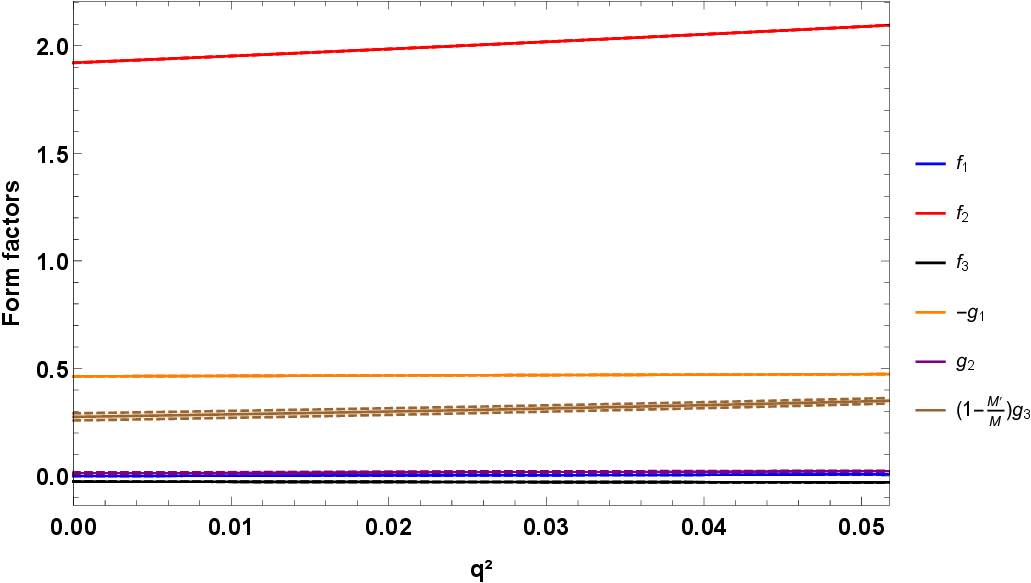}}
    \hfill
    \subfigure[\hspace{0.5em}$\Omega_{b}^{-}\to\Xi_{b}^{0}$]{\includegraphics[width=0.48\textwidth]{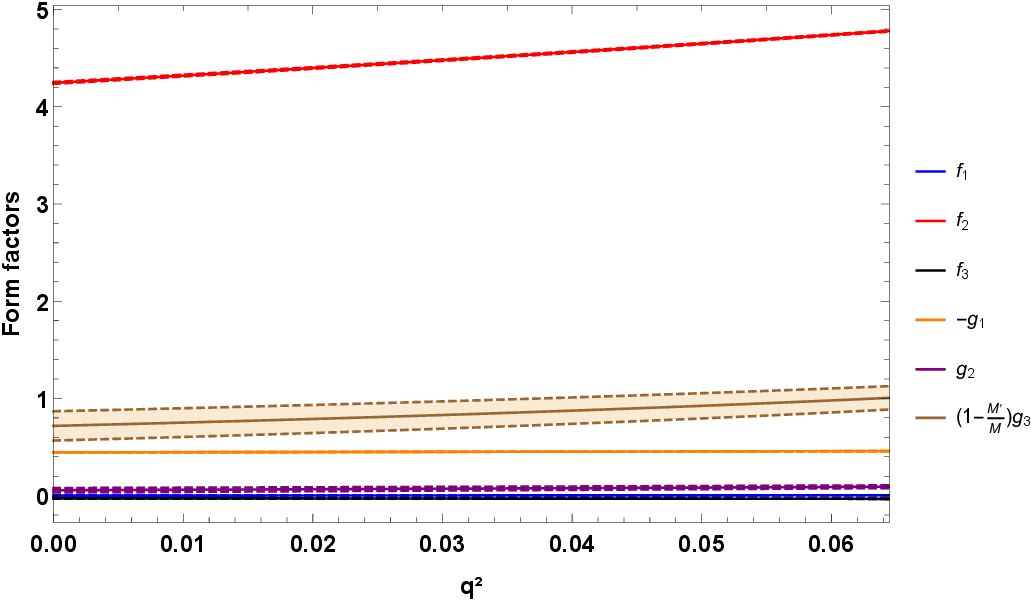}}

    \centering
    \subfigure[\hspace{0.5em}$\Omega_{c}^{0}\to\Xi_{c}^{\prime+}$]{\includegraphics[width=0.48\textwidth]{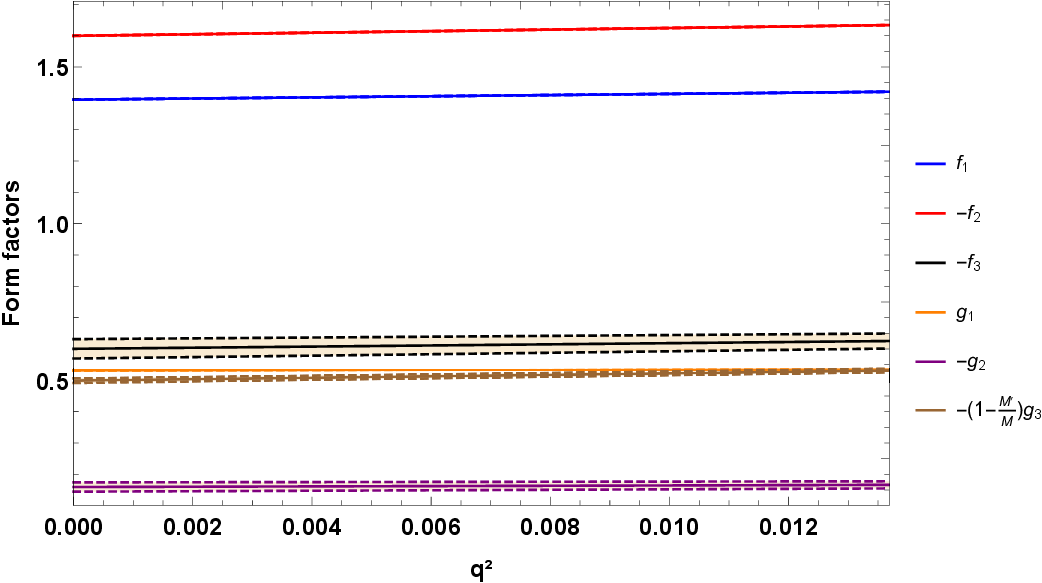}}
    \hfill
    \subfigure[\hspace{0.5em}$\Omega_{b}^{-}\to\Xi_{b}^{\prime0}$]{\includegraphics[width=0.48\textwidth]{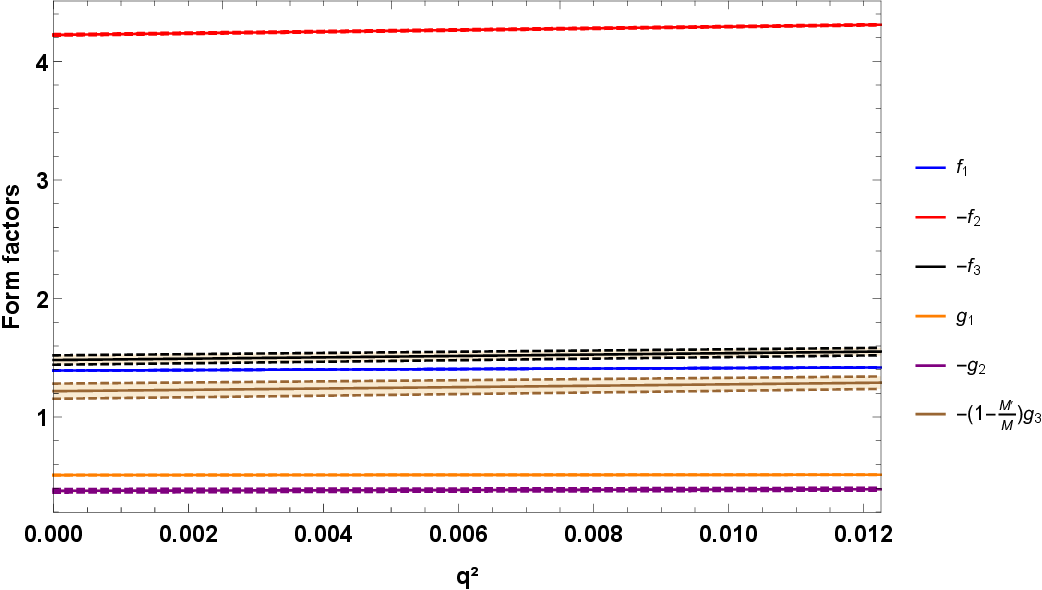}}
\caption{
The dependence of our form factors on $q^{2}$, where $g_{3}$ have been rescaled by a small factor ($1-M^{\prime}/M$) with $M^{(\prime)}$ the mass of the initial (final) baryon.
}
\label{fig:ff_q2}
\end{figure}

The fitted results of $a$ and $b$ are given in Table \ref{Tab:figi},
and the dependence of our form factors on $q^{2}$ can be found in
Fig. \ref{fig:ff_q2}. Some comments are in order. 
\begin{itemize}
\item The form factors of $\Xi_{c}^{+}\to\Sigma_{c}^{++}$, which are $\sqrt{2}$
times those of $\Xi_{c}^{0}\to\Sigma_{c}^{+}$, are not listed in
Table \ref{Tab:figi}. Here, $\sqrt{2}$ is the overlap factor, see
the discussion given at the beginning of Subsec. \ref{subsec:FF}. 
\item In Table \ref{Tab:figi}, we have considered all sources of error
mentioned above. However, one can see that these sources of error
only have a little impact on the form factors. Simply put, this is
because the transition matrix element $\langle{\cal B}_{f}(v^{\prime},s^{\prime})|\bar{u}\gamma_{\mu}(1-\gamma_{5})s|{\cal B}_{i}(v,s)\rangle$
is actually very similar to the baryon normalization inner product
$\langle{\cal B}(v^{\prime},s^{\prime})|{\cal B}(v,s)\rangle$, up
to some dynamic factors; In particular, the momentum wavefunctions
of the initial and final baryons are almost identical. By the way,
this is also the way we discuss the heavy quark limit for the form
factors.
\end{itemize}

\begin{table}
\caption{
Our fitted results of $a$ and $b$ for the form factors $f_{i}$ and $g_{i}$.
The form factors of $\Xi_{c}^{+}\to\Sigma_{c}^{++}$,
which are $\sqrt{2}$ times those of $\Xi_{c}^{0}\to\Sigma_{c}^{+}$,
are not shown here. 
}
\label{Tab:figi}%
\begin{tabular}{l|c|c|c}
\hline 
Transition & $(a,b)$ of $f_{1}$ & $(a,b)$ of $f_{2}$ & $(a,b)$ of $f_{3}$\tabularnewline
\hline 
$\Xi_{c}^{0}\to\Lambda_{c}^{+}$ & $(0.981\pm0.000,-0.798\pm0.004)$ & $(0.428\pm0.000,-0.409\pm0.004)$ & $(-0.185\pm0.006,0.696\pm0.154)$\tabularnewline
$\Xi_{b}^{-}\to\Lambda_{b}^{0}$ & $(0.983\pm0.000,-1.065\pm0.012)$ & $(0.458\pm0.000,-0.529\pm0.003)$ & $(-0.711\pm0.016,3.363\pm0.409)$\tabularnewline
\hline 
$\Xi_{c}^{0}\to\Sigma_{c}^{+}$ & $(-0.000\pm0.000,-0.153\pm0.001)$ & $(1.311\pm0.000,-0.935\pm0.015)$ & $(-0.019\pm0.000,0.085\pm0.001)$\tabularnewline
\hline 
$\Omega_{c}^{0}\to\Xi_{c}^{+}$ & $(0.007\pm0.000,-0.196\pm0.001)$ & $(1.959\pm0.000,-1.107\pm0.007)$ & $(-0.028\pm0.000,0.032\pm0.001)$\tabularnewline
$\Omega_{b}^{-}\to\Xi_{b}^{0}$ & $(0.004\pm0.000,-0.091\pm0.000)$ & $(4.392\pm0.005,-3.356\pm0.090)$ & $(-0.030\pm0.000,0.052\pm0.002)$\tabularnewline
\hline 
$\Omega_{c}^{0}\to\Xi_{c}^{\prime+}$ & $(1.397\pm0.000,-0.090\pm0.004)$ & $(-1.606\pm0.000,0.707\pm0.009)$ & $(-0.613\pm0.024,1.493\pm0.852)$\tabularnewline
$\Omega_{b}^{-}\to\Xi_{b}^{\prime0}$ & $(1.396\pm0.000,-0.552\pm0.005)$ & $(-4.243\pm0.005,2.534\pm0.184)$ & $(-1.526\pm0.031,5.789\pm1.131)$\tabularnewline
\hline 
Transition & $(a,b)$ of $g_{1}$ & $(a,b)$ of $g_{2}$ & $(a,b)$ of $g_{3}$\tabularnewline
\hline 
$\Xi_{c}^{0}\to\Lambda_{c}^{+}$ & $(-0.000\pm0.000,0.006\pm0.000)$ & $(0.001\pm0.000,-0.001\pm0.000)$ & $(-0.022\pm0.000,-0.111\pm0.003)$\tabularnewline
$\Xi_{b}^{-}\to\Lambda_{b}^{0}$ & $(-0.000\pm0.000,0.001\pm0.000)$ & $(0.000\pm0.000,-0.001\pm0.000)$ & $(-0.025\pm0.000,-0.090\pm0.009)$\tabularnewline
\hline 
$\Xi_{c}^{0}\to\Sigma_{c}^{+}$ & $(-0.345\pm0.000,-0.064\pm0.002)$ & $(0.153\pm0.002,-0.292\pm0.100)$ & $(53.789\pm1.480,-155.769\pm75.860)$\tabularnewline
\hline 
$\Omega_{c}^{0}\to\Xi_{c}^{+}$ & $(-0.458\pm0.000,-0.154\pm0.001)$ & $(0.021\pm0.003,-0.240\pm0.051)$ & $(3.263\pm0.119,-0.270\pm2.270)$\tabularnewline
$\Omega_{b}^{-}\to\Xi_{b}^{0}$ & $(-0.439\pm0.000,-0.124\pm0.006)$ & $(0.088\pm0.013,-0.747\pm0.214)$ & $(17.615\pm2.105,-12.249\pm33.873)$\tabularnewline
\hline 
$\Omega_{c}^{0}\to\Xi_{c}^{\prime+}$ & $(0.529\pm0.000,0.248\pm0.002)$ & $(-0.165\pm0.011,0.634\pm0.409)$ & $(-11.547\pm0.129,6.284\pm4.643)$\tabularnewline
$\Omega_{b}^{-}\to\Xi_{b}^{\prime0}$ & $(0.509\pm0.000,0.198\pm0.018)$ & $(-0.389\pm0.012,1.373\pm0.427)$ & $(-66.914\pm2.722,48.731\pm100.281)$\tabularnewline
\hline 
\end{tabular}
\end{table}

\subsection{Semileptonic decays}

The helicity amplitudes are defined by $H_{\lambda^{\prime},\lambda_{W}}^{V(A)}\equiv\langle{\cal B}^{\prime}(\lambda^{\prime})|\bar{u}\gamma^{\mu}(\gamma_{5})s|{\cal B}(\lambda)\rangle\epsilon_{W\mu}^{*}(\lambda_{W})$
with $\lambda=\lambda_{W}-\lambda^{\prime}$, which can be expressed
in terms of form factors: 
\begin{align}
 & H_{\frac{1}{2},1}^{V}=-i\sqrt{2Q_{-}}\left(f_{1}-\frac{M+M^{\prime}}{M}f_{2}\right),\quad H_{\frac{1}{2},0}^{V}=-i\frac{\sqrt{Q_{-}}}{\sqrt{q^{2}}}\left((M+M^{\prime})f_{1}-\frac{q^{2}}{M}f_{2}\right),\nonumber \\
 & H_{\frac{1}{2},t}^{V}=-i\frac{\sqrt{Q_{+}}}{\sqrt{q^{2}}}\left((M-M^{\prime})f_{1}+\frac{q^{2}}{M}f_{3}\right),\nonumber \\
 & H_{\frac{1}{2},1}^{A}=-i\sqrt{2Q_{+}}\left(g_{1}+\frac{M-M^{\prime}}{M}g_{2}\right),\quad H_{\frac{1}{2},0}^{A}=-i\frac{\sqrt{Q_{+}}}{\sqrt{q^{2}}}\left((M-M^{\prime})g_{1}+\frac{q^{2}}{M}g_{2}\right),\nonumber \\
 & H_{\frac{1}{2},t}^{A}=-i\frac{\sqrt{Q_{-}}}{\sqrt{q^{2}}}\left((M+M^{\prime})g_{1}-\frac{q^{2}}{M}g_{3}\right),\label{eq:HA}
\end{align}
where $Q_{\pm}=(M\pm M^{\prime})^{2}-q^{2}$, and $M^{(\prime)}$
is the mass of the initial (final) baryon. The other helicity amplitudes
can be obtained by 
\begin{equation}
H_{-\lambda^{\prime},-\lambda_{W}}^{V,A}=\pm H_{\lambda^{\prime},\lambda_{W}}^{V,A},
\end{equation}
and the total helicity amplitudes are
\begin{equation}
H_{\lambda^{\prime},\lambda_{W}}\equiv H_{\lambda^{\prime},\lambda_{W}}^{V}-H_{\lambda^{\prime},\lambda_{W}}^{A}.
\end{equation}

The polarized differential decay widths for ${\cal B}(1/2^{+})\to{\cal B}^{\prime}(1/2^{+})l\nu$
are 
\begin{align}
\frac{d\Gamma_{L}}{dq^{2}} & =\frac{G_{F}^{2}|V_{{\rm CKM}}|^{2}|\vec{P}^{\prime}|q^{2}(1-\hat{m}_{l}^{2})^{2}}{384\pi^{3}M^{2}}\left[(2+\hat{m}_{l}^{2})(|H_{\frac{1}{2},0}|^{2}+|H_{-\frac{1}{2},0}|^{2})+3\hat{m}_{l}^{2}(|H_{\frac{1}{2},t}|^{2}+|H_{-\frac{1}{2},t}|^{2})\right],\nonumber \\
\frac{d\Gamma_{T}}{dq^{2}} & =\frac{G_{F}^{2}|V_{{\rm CKM}}|^{2}|\vec{P}^{\prime}|q^{2}(1-\hat{m}_{l}^{2})^{2}}{384\pi^{3}M^{2}}(2+\hat{m}_{l}^{2})(|H_{\frac{1}{2},1}|^{2}+|H_{-\frac{1}{2},-1}|^{2}),\label{eq:dGammadq2}
\end{align}
where $\hat{m}_{l}\equiv m_{l}/\sqrt{q^{2}}$, and $|\vec{P}^{\prime}|$
is the magnitude of 3-momentum of ${\cal B}^{\prime}$ in the rest
frame of ${\cal B}$. The longitudinal polarization asymmetry $P_{L}$
is given as \cite{Ke:2019smy}
\begin{equation}
\resizebox{\textwidth}{!}{$\displaystyle
P_{L}=\frac{\int_{m_{l}^{2}}^{(M-M^{\prime})^{2}}dq^{2}\,|\vec{P}^{\prime}|q^{2}(1-\hat{m}_{l}^{2})^{2}\left[(2+\hat{m}_{l}^{2})(|H_{\frac{1}{2},0}|^{2}-|H_{-\frac{1}{2},0}|^{2}+|H_{\frac{1}{2},1}|^{2}-|H_{-\frac{1}{2},-1}|^{2})+3\hat{m}_{l}^{2}(|H_{\frac{1}{2},t}|^{2}-|H_{-\frac{1}{2},t}|^{2})\right]}{\int_{m_{l}^{2}}^{(M-M^{\prime})^{2}}dq^{2}\,|\vec{P}^{\prime}|q^{2}(1-\hat{m}_{l}^{2})^{2}\left[(2+\hat{m}_{l}^{2})(|H_{\frac{1}{2},0}|^{2}+|H_{-\frac{1}{2},0}|^{2}+|H_{\frac{1}{2},1}|^{2}+|H_{-\frac{1}{2},-1}|^{2})+3\hat{m}_{l}^{2}(|H_{\frac{1}{2},t}|^{2}+|H_{-\frac{1}{2},t}|^{2})\right]}.
$}
\end{equation}

\begin{table}
\caption{Our predictions for the semileptonic decays, where the first and second uncertainties are respectively from those of the form factors and the masses of initial and final baryons.}
\label{Tab:semi}%
\resizebox{\textwidth}{!}{
\begin{tabular}{l|c|c|c|c}
\hline 
Channel & $\Gamma/\text{~GeV}$ & ${\cal B}$ & $\Gamma_{L}/\Gamma_{T}$ & $P_{L}/\text{~GeV}$\tabularnewline
\hline 
$\Xi_{c}^{0}\to\Lambda_{c}^{+}e^{-}\bar{\nu}_{e}$ & $(6.68\pm0.00\pm0.08)\times10^{-19}$ & $(1.53\pm0.00\pm0.02)\times10^{-7}$ & $(4.46\pm0.02\pm0.05)\times10^{4}$ & $(6.02\pm0.12\pm0.29)\times10^{-5}$\tabularnewline
$\Xi_{b}^{-}\to\Lambda_{b}^{0}e^{-}\bar{\nu}_{e}$ & $(5.94\pm0.01\pm0.14)\times10^{-19}$ & $(1.42\pm0.00\pm0.03)\times10^{-6}$ & $(1.04\pm0.01\pm0.03)\times10^{6}$ & $(9.65\pm0.04\pm0.47)\times10^{-6}$\tabularnewline
\hline 
$\Xi_{c}^{+}\to\Sigma_{c}^{++}e^{-}\bar{\nu}_{e}$ & $(1.27\pm0.00\pm0.18)\times10^{-24}$ & $(0.87\pm0.00\pm0.12)\times10^{-12}$ & $1.25\pm0.00\pm0.00$ & $(-1.18\pm0.00\pm0.03)\times10^{-2}$\tabularnewline
$\Xi_{c}^{0}\to\Sigma_{c}^{+}e^{-}\bar{\nu}_{e}$ & $(2.31\pm0.00\pm0.31)\times10^{-24}$ & $(5.27\pm0.00\pm0.70)\times10^{-13}$ & $1.25\pm0.00\pm0.00$ & $(-1.53\pm0.00\pm0.05)\times10^{-2}$\tabularnewline
\hline 
$\Omega_{c}^{0}\to\Xi_{c}^{+}e^{-}\bar{\nu}_{e}$ & $(1.33\pm0.00\pm0.06)\times10^{-18}$ & $(5.50\pm0.00\pm0.24)\times10^{-7}$ & $1.18\pm0.00\pm0.00$ & $-0.20\pm0.00\pm0.00$\tabularnewline
$\Omega_{b}^{-}\to\Xi_{b}^{0}e^{-}\bar{\nu}_{e}$ & $(2.29\pm0.01\pm0.07)\times10^{-18}$ & $(5.70\pm0.01\pm0.17)\times10^{-6}$ & $1.16\pm0.00\pm0.00$ & $-0.23\pm0.00\pm0.00$\tabularnewline
\hline 
$\Omega_{c}^{0}\to\Xi_{c}^{\prime+}e^{-}\bar{\nu}_{e}$ & $(2.13\pm0.00\pm0.21)\times10^{-19}$ & $(8.86\pm0.00\pm0.86)\times10^{-8}$ & $6.51\pm0.01\pm0.02$ & $-0.69\pm0.00\pm0.00$\tabularnewline
$\Omega_{b}^{-}\to\Xi_{b}^{\prime0}e^{-}\bar{\nu}_{e}$ & $(1.64\pm0.00\pm0.10)\times10^{-19}$ & $(4.08\pm0.00\pm0.25)\times10^{-7}$ & $6.93\pm0.01\pm0.02$ & $-0.68\pm0.00\pm0.00$\tabularnewline
\hline 
\hline 
$\Xi_{c}^{0}\to\Lambda_{c}^{+}\mu^{-}\bar{\nu}_{\mu}$ & $(1.28\pm0.00\pm0.03)\times10^{-19}$ & $(2.93\pm0.00\pm0.06)\times10^{-8}$ & $(6.17\pm0.02\pm0.08)\times10^{4}$ & $(8.52\pm0.08\pm0.20)\times10^{-5}$\tabularnewline
$\Xi_{b}^{-}\to\Lambda_{b}^{0}\mu^{-}\bar{\nu}_{\mu}$ & $(0.98\pm0.00\pm0.04)\times10^{-19}$ & $(2.34\pm0.00\pm0.10)\times10^{-7}$ & $(1.48\pm0.02\pm0.05)\times10^{6}$ & $(1.39\pm0.01\pm0.04)\times10^{-5}$\tabularnewline
\hline 
$\Omega_{c}^{0}\to\Xi_{c}^{+}\mu^{-}\bar{\nu}_{\mu}$ & $(4.63\pm0.00\pm0.30)\times10^{-19}$ & $(1.92\pm0.00\pm0.12)\times10^{-7}$ & $0.90\pm0.00\pm0.00$ & $-0.20\pm0.00\pm0.00$\tabularnewline
$\Omega_{b}^{-}\to\Xi_{b}^{0}\mu^{-}\bar{\nu}_{\mu}$ & $(0.99\pm0.00\pm0.04)\times10^{-18}$ & $(2.48\pm0.01\pm0.10)\times10^{-6}$ & $0.93\pm0.00\pm0.00$ & $-0.23\pm0.00\pm0.00$\tabularnewline
\hline 
$\Omega_{c}^{0}\to\Xi_{c}^{\prime+}\mu^{-}\bar{\nu}_{\mu}$ & $(3.13\pm0.00\pm2.85)\times10^{-22}$ & $(1.30\pm0.00\pm1.18)\times10^{-10}$ & $4.22\pm0.01\pm0.05$ & $-0.27\pm0.00\pm0.02$\tabularnewline
$\Omega_{b}^{-}\to\Xi_{b}^{\prime0}\mu^{-}\bar{\nu}_{\mu}$ & $(1.57\pm0.00\pm2.15)\times10^{-23}$ & $(3.92\pm0.00\pm5.36)\times10^{-11}$ & $4.38\pm0.01\pm0.04$ & $-0.18\pm0.00\pm0.02$\tabularnewline
\hline 
\end{tabular}
}
\end{table}

Our predictions for the semileptonic decays are given in Table \ref{Tab:semi}.
When arriving at these results, we also have used \cite{ParticleDataGroup:2022pth}:
\begin{equation}
G_{F}=1.166\times10^{-5}\ {\rm GeV}^{-2},\quad|V_{us}|=0.225.
\end{equation}
Some comments are in order. 
\begin{itemize}
\item Due to the limitation of phase space, two semi-muonic decays are not
allowed. 
\item Most decay branching ratios are in the range of $10^{-8}$-$10^{-6}$,
while those for $\Xi_{c}^{+}\to\Sigma_{c}^{++}e^{-}\bar{\nu}_{e}$,
$\Xi_{c}^{0}\to\Sigma_{c}^{+}e^{-}\bar{\nu}_{e}$ and $\Omega_{c}^{0}\to\Xi_{c}^{\prime+}\mu^{-}\bar{\nu}_{\mu}$,
$\Omega_{b}^{-}\to\Xi_{b}^{\prime0}\mu^{-}\bar{\nu}_{\mu}$ are even
smaller due to their extremely small phase space. 
\item For the $0^{+}\to0^{+}$ process, $\Gamma_{L}$ has an overwhelming
advantage over $\Gamma_{T}$. For the $0^{+}\to1^{+}$ and $1^{+}\to0^{+}$
processes, $\Gamma_{L}$ is comparable with $\Gamma_{T}$. For the
$1^{+}\to1^{+}$ process, $\Gamma_{L}$ is several times larger
than $\Gamma_{T}$. 
\item It is particularly interesting to consider the lepton flavor universality
(LFU) for semileptonic $s\to u$ decays of singly heavy baryons. Our predictions are given below:
\begin{align}
 & \frac{\Gamma(\Xi_{c}^{0}\to\Lambda_{c}^{+}\mu^{-}\bar{\nu}_{\mu})}{\Gamma(\Xi_{c}^{0}\to\Lambda_{c}^{+}e^{-}\bar{\nu}_{e})}=0.19\pm0.00\pm0.00,\quad\frac{\Gamma(\Xi_{b}^{-}\to\Lambda_{b}^{0}\mu^{-}\bar{\nu}_{\mu})}{\Gamma(\Xi_{b}^{-}\to\Lambda_{b}^{0}e^{-}\bar{\nu}_{e})}=0.16\pm0.00\pm0.00,\nonumber \\
 & \frac{\Gamma(\Omega_{c}^{0}\to\Xi_{c}^{+}\mu^{-}\bar{\nu}_{\mu})}{\Gamma(\Omega_{c}^{0}\to\Xi_{c}^{+}e^{-}\bar{\nu}_{e})}=0.35\pm0.00\pm0.01,\quad\frac{\Gamma(\Omega_{b}^{-}\to\Xi_{b}^{0}\mu^{-}\bar{\nu}_{\mu})}{\Gamma(\Omega_{b}^{-}\to\Xi_{b}^{0}e^{-}\bar{\nu}_{e})}=0.43\pm0.00\pm0.01,\nonumber \\
 & \frac{\Gamma(\Omega_{c}^{0}\to\Xi_{c}^{\prime+}\mu^{-}\bar{\nu}_{\mu})}{\Gamma(\Omega_{c}^{0}\to\Xi_{c}^{\prime+}e^{-}\bar{\nu}_{e})}=(1.47\pm0.00\pm1.09)\times10^{-3},\nonumber \\
 & \frac{\Gamma(\Omega_{b}^{-}\to\Xi_{b}^{\prime0}\mu^{-}\bar{\nu}_{\mu})}{\Gamma(\Omega_{b}^{-}\to\Xi_{b}^{\prime0}e^{-}\bar{\nu}_{e})}=(0.96\pm0.00\pm1.18)\times10^{-4},\label{eq:R_values}
\end{align}
where the first and second uncertainties are respectively from those of the form factors and the masses of initial and final baryons. 
One can see that, the decay width of a semi-muonic decay is significantly
smaller than that of the corresponding semi-electronic decay due to
the smaller phase space, and the last two ratios are particularly
small. 
\item Consider the heavy quark limit for semi-electronic decays, where the
electron mass can be neglected. Starting from the complete expressions
of the decay widths in Eq.~(\ref{eq:dGammadq2}), one can arrive at
$\Gamma=\frac{G_{F}^{2}|V_{{\rm CKM}}|^{2}(M-M^{\prime})^{5}}{60\pi^{3}}(f_{1}^{2}+3g_{1}^{2})$
in the heavy quark limit. When substituting the heavy quark limit
of the form factors in Table \ref{Tab:ff_HQL} into this equation,
we further obtain $\Gamma=\frac{G_{F}^{2}|V_{{\rm CKM}}|^{2}(M-M^{\prime})^{5}}{60\pi^{3}}$
for the three processes of $0^{+}\to0^{+}$, $0^{+}\to1^{+}$ and
$1^{+}\to0^{+}$, which are consistent with the conclusions in Ref.
\cite{Faller:2015oma}. However, for the $1^{+}\to1^{+}$ process, we have $\Gamma=\frac{G_{F}^{2}|V_{{\rm CKM}}|^{2}(M-M^{\prime})^{5}}{60\pi^{3}}\times\frac{7}{3}$,
which is smaller than the result $\Gamma=\frac{G_{F}^{2}|V_{{\rm CKM}}|^{2}(M-M^{\prime})^{5}}{15\pi^{3}}$
in Ref. \cite{Faller:2015oma}. There are two points worth noting here. First,
our results here do not include the overlap factors, see the caption
of Table \ref{Tab:ff_HQL}. Second, as mentioned earlier, Ref. \cite{Faller:2015oma}
considers the heavy quark limit of form factors at $q^{2}=q_{{\rm max}}^{2}$,
while we consider it at $q^{2}=0$.
\item The heavy quark symmetry breaking can be quantitatively studied. For
example, the breaking degree between $\Xi_{c}^{0}\to\Lambda_{c}^{+}e^{-}\bar{\nu}_{e}$
and $\Xi_{b}^{-}\to\Lambda_{b}^{0}e^{-}\bar{\nu}_{e}$ is about 10\%. 
\item In Table \ref{Tab:semi}, we have respectively considered the uncertainties
from the form factors and the masses of initial and final baryons.
One can see that, the uncertainty from the former is small, and can
even be neglected. This is because the decay width is only significantly
dependent on $f_{1}$ and $g_{1}$, and their errors are all very
small, as can be seen in Table \ref{Tab:figi}. Somewhat unexpectedly,
the masses of the initial and final baryons are the main sources of
error for the decay width. Consider two extreme cases: $\Xi_{c}^{+}\to\Sigma_{c}^{++}e^{-}\bar{\nu}_{e}$
and $\Xi_{c}^{0}\to\Sigma_{c}^{+}e^{-}\bar{\nu}_{e}$.
Note that, if the SU(3) flavor symmetry strictly holds, the decay width of the
former should be twice that of the latter. 
However, in reality, the decay width of the latter
is almost twice that of the former! This is because
$m_{\Xi_{c}^{+}}-m_{\Sigma_{c}^{++}}\approx14$ MeV, $m_{\Xi_{c}^{0}}-m_{\Sigma_{c}^{+}}\approx18$
MeV, and $18^{5}$ is almost four times larger than $14^{5}$. 
\end{itemize}

\subsection{Comparison}

In Subsec. \ref{subsec:HQL}, we have compared our form factors in
the heavy quark limit with those in Ref. \cite{Faller:2015oma}. In
this subsection, we also compare our predicted semileptonic decay
widths with those in the literature, as can be seen in Table~\ref{Tab:comparison_semi}.
In this table, Ref. \cite{Faller:2015oma} considered the heavy quark
limit, Ref. \cite{Soni:2015gwl} employed the extended harmonic confinement
model, and the authors of Ref. \cite{Shah:2023qsg} used their spectral
parameters. Some comments are in order. 
\begin{itemize}
\item In Table \ref{Tab:comparison_semi}, we present the decay widths calculated
using the full form factors and decay width formula, while Refs. \cite{Faller:2015oma}, \cite{Soni:2015gwl},
and \cite{Shah:2023qsg} present the results in the heavy quark limit.
\item From Table \ref{Tab:comparison_semi}, it can be seen that the results
of Refs. \cite{Faller:2015oma}, \cite{Soni:2015gwl}, and \cite{Shah:2023qsg}
are basically consistent, except for $\Omega_{c}^{0}\to\Xi_{c}^{+}e^{-}\bar{\nu}_{e}$.
On the one hand, it is not surprising that their results are basically
consistent, as both Refs. \cite{Soni:2015gwl} and \cite{Shah:2023qsg}
use the decay width formula from Ref. \cite{Faller:2015oma}, which
is only proportional to $(M-M^{\prime})^{5}$ except for some constants.
On the other hand, for $\Omega_{c}^{0}\to\Xi_{c}^{+}e^{-}\bar{\nu}_{e}$,
the authors of Ref. \cite{Soni:2015gwl} mistakenly quoted the formula
from Ref. \cite{Faller:2015oma}, and the correct result should be
further divided by 4. 
\item In the last subsection, we also obtained the decay width formulas for the semi-electronic decays in the heavy quark limit.
One can see that, $f_{1}$ and $g_{1}$ play the main roles.
For the processes of $0^{+}\to0^{+}$, $0^{+}\to1^{+}$, and $1^{+}\to0^{+}$, our decay width formulas are consistent with those in Ref. \cite{Faller:2015oma}, while for the $1^{+}\to1^{+}$ process,  our decay width is smaller. When arriving at our decay widths involving $\Sigma_{c}^{++}$, $\Omega_{c}^{0}$, and $\Omega_{b}^{-}$, the squared overlap factor $(\sqrt{2})^{2}$ should be considered.
\item From Table \ref{Tab:comparison_semi}, it can be seen that for the $0^{+}\to0^{+}$ process, our results are
close to those in the heavy quark limit in Ref. \cite{Faller:2015oma};
For the $0^{+}\to1^{+}$ and $1^{+}\to0^{+}$
processes, our results are approximately 2-3 times smaller than those
in Ref. \cite{Faller:2015oma}; For the
$1^{+}\to1^{+}$ process, our results are expected to be smaller than
those in Ref. \cite{Faller:2015oma}, but they turn out to be somewhat too small.
All these differences can be explained by comparing our form factors in the heavy quark
limit in Table \ref{Tab:ff_HQL} and our calculated form factors in
Table \ref{Tab:f0g0_with_errs}. The main issue is that the calculated
$g_{1}$ in the processes of $0^{+}\to1^{+}$ and $1^{+}\to0^{+}$
are significantly smaller than those in the heavy quark limit. Tracing
back to the source of the problem, it should be that both the $s$-quark
and $u$-quark are taken as massless in the heavy quark limit, which
may have led to this apparent deviation. 
In other words, the $m_{s}$ correction may not be negligible.
\item In addition, in Ref. \cite{Faller:2015oma}, $m_{\Omega_{b}}=6048.8$
MeV, while in this work, $m_{\Omega_{b}}=6045.8\pm0.8$ MeV, which
results in an error of approximately 10\% for $\Omega_{b}^{-}\to\Xi_{b}^{\prime0}e^{-}\bar{\nu}_{e}$.
\end{itemize}

\begin{table}
\caption{Our predictions on the semileptonic decay widths (in units of GeV)
are compared with those in the literature. }
\label{Tab:comparison_semi} %
\begin{tabular}{l|c|c|c|c}
\hline 
Channel  & This work & Ref. \cite{Faller:2015oma}  & Ref. \cite{Soni:2015gwl}  & Ref. \cite{Shah:2023qsg}\tabularnewline
\hline 
$\Xi_{c}^{0}\to\Lambda_{c}^{+}e^{-}\bar{\nu}_{e}$  & $(6.68\pm0.00\pm0.08)\times10^{-19}$ & $7.91\times10^{-19}$  & - -  & $7.839\times10^{-19}$\tabularnewline
$\Xi_{b}^{-}\to\Lambda_{b}^{0}e^{-}\bar{\nu}_{e}$  & $(5.94\pm0.01\pm0.14)\times10^{-19}$ & $6.16\times10^{-19}$  & - -  & $5.928\times10^{-19}$\tabularnewline
\hline 
$\Xi_{c}^{+}\to\Sigma_{c}^{++}e^{-}\bar{\nu}_{e}$  & $(1.27\pm0.00\pm0.18)\times10^{-24}$ & $3.74\times10^{-24}$  & - -  & - -\tabularnewline
$\Xi_{c}^{0}\to\Sigma_{c}^{+}e^{-}\bar{\nu}_{e}$  & $(2.31\pm0.00\pm0.31)\times10^{-24}$ & $6.97\times10^{-24}$  & - -  & $7.023\times10^{-24}$\tabularnewline
\hline 
$\Omega_{c}^{0}\to\Xi_{c}^{+}e^{-}\bar{\nu}_{e}$  & $(1.33\pm0.00\pm0.06)\times10^{-18}$ & $2.26\times10^{-18}$  & $9.05\times10^{-18}$  & $2.290\times10^{-18}$\tabularnewline
$\Omega_{b}^{-}\to\Xi_{b}^{0}e^{-}\bar{\nu}_{e}$  & $(2.29\pm0.01\pm0.07)\times10^{-18}$ & $4.05\times10^{-18}$  & - -  & $4.007\times10^{-18}$\tabularnewline
\hline 
$\Omega_{c}^{0}\to\Xi_{c}^{\prime+}e^{-}\bar{\nu}_{e}$  & $(2.13\pm0.00\pm0.21)\times10^{-19}$ & $3.63\times10^{-19}$  & $3.65\times10^{-19}$  & - -\tabularnewline
$\Omega_{b}^{-}\to\Xi_{b}^{\prime0}e^{-}\bar{\nu}_{e}$  & $(1.64\pm0.00\pm0.10)\times10^{-19}$ & - -  & - -  & - -\tabularnewline
\hline 
\hline 
$\Xi_{c}^{0}\to\Lambda_{c}^{+}\mu^{-}\bar{\nu}_{\mu}$  & $(1.28\pm0.00\pm0.03)\times10^{-19}$ & $1.3\times10^{-19}$  & - -  & - -\tabularnewline
$\Xi_{b}^{-}\to\Lambda_{b}^{0}\mu^{-}\bar{\nu}_{\mu}$  & $(0.98\pm0.00\pm0.04)\times10^{-19}$ & $0.91\times10^{-19}$  & - -  & - -\tabularnewline
\hline 
$\Omega_{c}^{0}\to\Xi_{c}^{+}\mu^{-}\bar{\nu}_{\mu}$  & $(4.63\pm0.00\pm0.30)\times10^{-19}$ & $7.1\times10^{-19}$  & - -  & - -\tabularnewline
$\Omega_{b}^{-}\to\Xi_{b}^{0}\mu^{-}\bar{\nu}_{\mu}$  & $(0.99\pm0.00\pm0.04)\times10^{-18}$ & $1.7\times10^{-18}$  & - -  & - -\tabularnewline
\hline 
$\Omega_{c}^{0}\to\Xi_{c}^{\prime+}\mu^{-}\bar{\nu}_{\mu}$  & $(3.13\pm0.00\pm2.85)\times10^{-22}$ & $10\times10^{-22}$  & - -  & - -\tabularnewline
$\Omega_{b}^{-}\to\Xi_{b}^{\prime0}\mu^{-}\bar{\nu}_{\mu}$  & $(1.57\pm0.00\pm2.15)\times10^{-23}$ & - -  & - -  & - -\tabularnewline
\hline 
\end{tabular}
\end{table}

\section{Conclusions}

In this work, we investigate semileptonic
$s\to u$ decays of singly heavy baryons using the light-front approach
under the three-quark picture. Firstly, we provide all the form factors
in this quark model, especially for $f_{3}$ and $g_{3}$, which are
usually considered non-extractable in recent literature. We also discuss
the heavy quark limit for the form factors, and some of our results
differ from those in the literature. Secondly, we apply the obtained
form factors to give some predictions on the semileptonic decays.
We also compare our predicted decay widths with those in the heavy quark limit in the literature.
We find that, the decay width formulas in the heavy quark limit may deviate significantly from the full formula, due to the neglect of the $m_{s}$ correction.
One of the most unique aspects is the examination of lepton
flavor universality (LFU). Due to the extremely small phase space,
LFU is highly sensitive to the masses of leptons, and precise measurement
of LFU for these processes is expected to be an important tool for
testing the standard model. 

\section*{Acknowledgements}

The authors are grateful to Lei-Yi Li, Run-Hui Li, Wei Wang, Zhi-Peng
Xing, Yu-Ji Shi, Dan Zhang for valuable discussions. This work is
supported in part by National Natural Science Foundation of China
under Grant No.~12465018.

\end{document}